\documentclass[a4paper,11pt,fleqn]{article}

\usepackage{amsmath,amssymb,amsthm} 

\usepackage{hyperref}
\hypersetup{colorlinks, linkcolor=blue, citecolor=blue, urlcolor=blue}

\allowdisplaybreaks
\newcommand{\p}{\partial}
\newcommand{\const}{\mathop{\rm const}\nolimits}
\newcommand{\lsemioplus}{\mathbin{\mbox{$\lefteqn{\hspace{.77ex}\rule{.4pt}{1.2ex}}{\in}$}}}
\newcommand{\spanindex}{{\mbox{\tiny$\langle\,\rangle$}}}

\newcommand{\vv}{\mathbf{v}}
\newcommand{\xx}{\mathbf{x}}
\newcommand{\PP}{\mathcal{P}}
\newcommand{\SSS}{\mathcal{S}}
\newcommand{\ZZ}{\mathcal{Z}}
\newcommand{\XX}{\mathcal{X}}
\newcommand{\DD}{\mathcal{D}}
\newcommand{\JJ}{\mathcal{J}}

\newcommand{\todo}[1][\null]{\ensuremath{\clubsuit}}

\newcommand{\noprint}[1]{}

\newtheorem{theorem}{Theorem}

\newtheorem{corollary}{Corollary}

{\theoremstyle{definition}

\newtheorem{remark}{Remark}
\newtheorem*{remark*}{Remark}
}

\begin{document}

\par\noindent {\LARGE\bf
On the ineffectiveness of constant rotation \\ in the primitive equations and their symmetry analysis
\par}

{\vspace{5mm}\par\noindent {\bf Elsa Dos Santos Cardoso-Bihlo$^\dag$ and Roman O.\ Popovych$^\ddag$
} \par\vspace{3mm}\par}

{\vspace{2mm}\par\noindent {\it
$^{\dag}$Department of Mathematics and Statistics, Memorial University of Newfoundland,\\
$\phantom{^{\dag}}$~St.\ John's (NL) A1C 5S7, Canada
}

\vspace{2mm}\par\noindent{\it
$^\ddag$\,Fakult\"at f\"ur Mathematik, Universit\"at Wien, Oskar-Morgenstern-Platz 1, A-1090 Wien, Austria%
\\
$\phantom{^\ddag}$Institute of Mathematics of NAS of Ukraine, 3 Tereshchenkivska Str., 01024 Kyiv, Ukraine
}\par}

{\vspace{3mm}\par\noindent
$\phantom{^\dag}$~\textup{E-mail}:
$^{\dag}$ecardosobihl@mun.ca, $^{\ddag}$rop@imath.kiev.ua
\par}

\vspace{8mm}\par\noindent\hspace*{8mm}\parbox{140mm}{\small
Modern weather and climate prediction models are based on a system of nonlinear partial differential equations called the primitive equations. Lie symmetries of the primitive equations with zero external heating rate are computed and the structure of its maximal Lie invariance algebra, which is infinite-dimensional, is studied. The maximal Lie invariance algebra for the case of a nonzero constant Coriolis parameter is mapped to the case of vanishing Coriolis force. The same mapping allows one to transform the constantly rotating primitive equations to the equations in a resting reference frame. This mapping is used to obtain exact solutions for the rotating case from exact solutions for the nonrotating equations. Another important result of the paper is the computation of the complete point symmetry group of the primitive equations using the algebraic method.
}\par\vspace{3mm}

\noprint{

MSC: 76M60 (Primary) 76U60 86A10 35A30 35B06 35C05 (Secondary)
76-XX   Fluid mechanics {For general continuum mechanics, see 74Axx, or other parts of 74-XX} 
 76Mxx  Basic methods in fluid mechanics [See also 65-XX] 
  76M60   Symmetry analysis, Lie group and algebra methods 
  76Uxx Rotating fluids
   76U60  Geophysical flows [See also 86A05, 86A10]
86-XX  Geophysics [See also 76U05, 76V05]
  86Axx Geophysics [See also 76U05, 76V05]
   86A10  Meteorology and atmospheric physics [See also 76Bxx, 76E20, 76N15, 76Q05, 76Rxx, 76U05]
35-XX   Partial differential equations 
 35Axx  General topics 
  35A30   Geometric theory, characteristics, transformations [See also 58J70, 58J72] 
 35Bxx  Qualitative properties of solutions
  35B06   Symmetries, invariants, etc.  
 35Cxx  Representations of solutions 
  35C05   Solutions in closed form


}

\section{Introduction}

One of the main motivations for the study of Lie symmetries of partial differential equations is that they provide systematic tools which allow finding of ansatzes that reduce the number of independent variables. Depending on the particular form of reduction ansatzes, reduced systems of differential equations can then be often integrated to yield exact solutions, which gives particular solutions of the initial system of partial differential equations.
By their definition, Lie symmetries also can be used for generating new exact solutions from known ones.

Another important application of symmetries of differential equations is that they can provide a necessary condition of whether two equations can be mapped to each other. Thus, this criterion is effective in the case when the target equation is linear as then the initial equation is linearizable. For the equations of hydro-thermodynamics, Lie symmetries have proved to be extremely successful in finding point, contact and even nonlocal transformations relating different equations.
A number of equations, such as the one-dimensional shallow-water equations, the Thomas equation and the potential Burgers equation, are linearizable by point transformations~\cite{blum89Ay,blum10Ay,hydo00Ay,kume82Ay}.
Some equations are linearizable in a nonlocal way, e.g., by a point transformation after introducing potentials.
The most famous example for a nonlocal transformation is certainly the linearization of the Burgers equation by means of the Hopf--Cole transformation~\cite{olve86Ay},
which in fact is a non-invertible transformation, mapping the Burgers equation to the linear heat equation.
The $(-2)$-power diffusion equation and some nonlinear wave equations are also linearized by point transformations after potentialization as well.
Nonlinear differential equations can also be reduced by point transformations to other nonlinear differential equations of simpler form,
e.g., the cylindrical Korteweg--de Vries equation to the classical Korteweg--de Vries equation.
The Liouville equation is linearized by a differential substitution.
All the above transformations can be found by invoking the structure of the maximal Lie invariance algebras of the equations involved.
For invertible point transformations as will be considered in the present paper, the relevant necessary criterion for the existence of a mapping relating two system of differential equations to each other is that the maximal Lie invariance algebras of the initial and the target system are isomorphic~\cite{blum89Ay,blum10Ay}.

Quite recently, point transformations allowing for canceling terms related to the Coriolis force were found for a number of models of fluid dynamics. Although somehow expectable from the physical point of view, these transformations are often nontrivial. Examples of particular models where such a transformation was already found are the vorticity equation in spherical coordinates~\cite{bihl09Ay,bihl12Ay,plat60Ay}, the barotropic potential vorticity equation~\cite{bihl09Cy} and the shallow-water equations on flat~\cite{ches09Ay} and parabolic topography~\cite{ches11Ay}.

The purpose of the present paper is to show that a transformation eliminating the Coriolis force also exists for the more complex system of the primitive equations, which are nonlinear partial differential equations for the momentum, mass and energy conservation of atmospheric flows.
The primitive equations form the dynamical core of most of the modern large-scale weather and climate prediction models.

A further major result of the present paper is the computation of the complete point symmetry group of the primitive equations with vanishing external heating rate using the megaideal-based version of the algebraic method proposed in~\cite{bihl11Cy} and further developed in~\cite{bihl2015a,card12Ay,malt2021a}.
The method applied can be seen as a refinement of the general algebraic technique suggested in~\cite{hydo00By,hydo00Ay} (see also~\cite{gray13Ay}) by using the notion of megaideals (i.e., characteristically nilpotent ideals) of Lie algebras~\cite{popo03Ay}.
To the best of our knowledge, this is the first example of computing, within the framework of the algebraic approach, the complete point symmetry group for a multidimensional system of nonlinear partial differential equations whose maximal Lie invariance algebra is infinite-dimensional and is of complicated structure. The core part of the computation procedure is the construction of a sufficiently wide set of megaideals that place suitable restrictions on admitted point symmetries of the primitive equations. Without the application of the algebraic method, finding the complete point symmetry group would require the solution of a cumbersome nonlinear system of partial differential equations, which in general is a hopeless endeavor.

The further structure of this paper is as follows. In Section~\ref{sec:PrimitiveEquations}, the primitive equations are introduced. In Section~\ref{sec:SymmetriesPrimitiveEquations}, we compute the Lie symmetries of the primitive equations, where the external heating rate is zero, and explicitly find a point transformation that allows canceling of the effects of a constant rotation. In Section~\ref{sec:ompletePointSymmetryGroupPrimitiveEquations}, we determine the complete point symmetry group of the primitive equations with zero external heating rate in a resting reference frame using the algebraic method. Section~\ref{sec:SolutionsPrimitiveEquations} is devoted to the usage of the transformation found and the computation of selected exact solutions of the same primitive equations. 
In Section~\ref{sec:OnEquivTrans}, we construct the usual and generalized equivalence algebras of the class of the primitive equations. 
We also discuss the optimal way for solving the group classification problem for this class, which is based on the algebraic method of group classification. 
The final Section~\ref{sec:ConclusionPrimitiveEquations} briefly sums up the results of the paper.
\looseness=-1

\section{The primitive equations}\label{sec:PrimitiveEquations}

The frictionless primitive equations in the standard Cartesian coordinates are
\begin{gather}\label{eq:PrimitiveEquationsInCartesianCoords}
\begin{split}
 &\vv_t +\vv\cdot\nabla\vv + w\vv_z+f(-v,u)^{\rm T} + \frac{1}{\rho}\nabla p=0,\\
 &p_z + g\rho = 0,\\[1.5ex]
 &\rho_t + \nabla\cdot(\rho\vv) + (\rho w)_z= 0,\\
 &T_t+\vv\cdot\nabla T + w T_z - \frac {R}{c_p}\frac{T}{p}\frac{\mathrm{d}p}{\mathrm{d}t} = \frac{J}{c_p},
\end{split}
\end{gather}
where $\vv=(u,v)$ is the horizontal component of the velocity vector, $\nabla=(\p_x,\p_y)$ is the two-dimensional nabla operator, $w$~is the vertical velocity in the Cartesian coordinate system, $p$~is the pressure and $T$~is the temperature. All the unknown functions, $\vv$, $w$, $p$ and $T$ depend on $(t,x,y,z)$. Subscripts of functions denote differentiation with respect to the corresponding variables. The constants~$f$, $g$, $R$, $c_p$ in the above system are the Coriolis parameter, the free fall acceleration, the gas constant for dry air and the specific heat of dry air at constant pressure. The function $J=J(t,x,y,z)$ is the external heating. The equations in the system~\eqref{eq:PrimitiveEquationsInCartesianCoords} are respectively the momentum equation, the hydrostatic equation, the continuity equation and a version of the first law of hydrodynamics. 
In addition, we have that $g=\phi_z$, with $\phi$ being the geopotential. 
The density is related to the temperature and the pressure via the state equation of an ideal gas, $\rho=p/(RT)$.

Due to modern weather prediction models being formulated using the pressure (or a suitable function of the pressure) as the vertical coordinate rather than the height $z$ itself, it is convenient to work with the following form of the primitive equations instead: 
\begin{gather}\label{eq:PrimitiveEquations}
\begin{split}
 &\vv_t +\vv\cdot\nabla\vv +\omega \vv_p+f(-v,u)^{\rm T} + \nabla\phi=0,\\[.5ex]
 &\phi_p + \frac RpT = 0,\\[.5ex]
 &u_x+v_y+\omega_p = 0,\\[.5ex]
 &T_t+\vv\cdot\nabla T + \omega T_p - \frac R{c_p}\frac\omega pT = \frac{J}{c_p}.
\end{split}
\end{gather}
This form can be derived from the system~\eqref{eq:PrimitiveEquationsInCartesianCoords} using 
$p$ instead of~$z$ as the new vertical coordinate, 
the material derivative $\mathrm{d}p/\mathrm{d}t=:\omega$ of the pressure~$p$ instead of~$w$ as the vertical velocity in the pressure coordinate system, and
the geopotential~$\phi$ instead of~$p$ as one more new unknown function, 
see~\cite{kasa74Ay} for further details. 
Here, all the unknown functions, $\vv$, $\omega$, $\phi$ and $T$, now depend on $(t,x,y,p)$. 
From the physical point of view, the system~\eqref{eq:PrimitiveEquations} forms the dynamical core of most of the present day's atmospheric numerical models.

The physical constants~$R$ and~$c_p$ are always positive. Moreover, from the practical point of view, the thermodynamic relation $c_p=c_v+R$ applies for ideal gases. By definition, $c_v$ is the specific heat at constant volume, which is the amount of energy needed to heat one kilogram of a compound by one Kelvin while holding the volume constant. As there is no compound which will be heated by one Kelvin without supplying energy (i.e., always $c_v>0$), this implies that $c_p>R$. So as to simplify the subsequent expressions, we denote $\kappa:=R/c_p$, and thus $0<\kappa<1$. For the Coriolis parameter~$f$ we distinguish between the cases of $f=0$ (no rotation of the reference frame) and $f=\const$ (constant rotation of the reference frame). There arises the question of whether the choice $f=\const$ is a physically interesting one. By definition, $f=2\Omega\sin\varphi$, where $\Omega$ is the angular velocity of the Earth and $\varphi$ is the geographic latitude. Thus, $f=f(\varphi)$ and therefore changes along the meridians. On the other hand, this change is rather small and eventually can be neglected for domains extending only moderately in North--South direction. For example, for a domain extending approximately $300$ kilometer in North--South direction, the relative change in the value of $f$ from the South to the North is only about 5\% around the mid-latitudes. Therefore, for processes that take place on relatively small domains, $f=\const$ is a good approximation.

A process that can be described with the model~\eqref{eq:PrimitiveEquations} for $f=\const$ is the land--sea breeze. This is a circulation often induced by differential heating of a land-sea boundary, with winds directed landward during day and seaward during night.  As the land--sea breeze can persist for several hours, the effect of the Coriolis force cannot be neglected. It is generally found that around six hours after the beginning of the sea breeze the circulation is weakened due to the effects of the Coriolis force~\cite{cros10Ay}. This is why $f=\const$ is essential in numerical models that aim to capture the land--sea circulation in an accurate way, see~\cite{cros10Ay,phys76Ay} and references therein for a detailed review over numerical studies of this particular circulation pattern.

Another reason why it is convenient to assume $f=\const$ in the above system is the usage of Cartesian coordinates. For processes taking place on a large enough domain, the tangential plane approximation of the Earth is not reasonable any more. For such processes or for the general description of the global atmospheric circulation, it is more appropriate to study the primitive equations in spherical coordinates and to use the equality $f=2\Omega\sin\varphi$ without approximation.

Within the framework of group analysis of differential equations, the parameterized system~\eqref{eq:PrimitiveEquations} 
should be interpreted as a class of systems of differential equations with the arbitrary elements $f$, $R$, $c_p$ and $J$. 
An alternative option is to denote $\kappa:=R/c_p$ and $\hat J:=J/c_p$ and use $(f,R,\kappa,\hat J)$ instead of $(f,R,c_p,J)$ 
as arbitrary-element tuple for the class~\eqref{eq:PrimitiveEquations}.
Two of the arbitrary elements are inessential, in that it is possible to scale 
$R=1$ (by a scaling of $T$, $R$, $c_p$ and~$J$) and $f=1$ if $f\ne0$ (by a scaling of $(t,x,y,p)$, $f$, $c_p$ and~$J$). 
For physical reasons we will not make a use of these scalings. 
Moreover, the possibility to set~$f=1$ is not overly relevant as we will show in Section~\ref{sec:SymmetriesPrimitiveEquations} 
that $f$ can be set to zero by a point transformation.

In what follows we will mostly be concerned with the system~\eqref{eq:PrimitiveEquations} in the case $J=0$ corresponding to a non-heated atmosphere (the isentropic case).

\section{Lie symmetries}\label{sec:SymmetriesPrimitiveEquations}

We now compute Lie symmetries for the isentropic case when the external heating rate vanishes, i.e., $J=0$. The reason for considering this case specifically is twofold. Firstly, physically speaking the value $J$ constitutes the driver of the primitive equations, which includes various processes such as short- and long-wave radiation, friction, boundary layer effects and other, see e.g.~\cite{chri91Ay}, which normally have to be parameterized. As such, no convenient closed-form expression exists for this term. Secondly, and related to the first point, as $J$ has to be parameterized, within the framework of invariant parameterization put forth in~\cite{bihl12Dy, bihl11Fy, popo10Cy} it is convenient to consider the unparameterized model corresponding to $J=0$ and to compute all relevant group-theoretical objects, such as symmetries and conservation laws, for this case first, which then allows one to find suitable expressions for $J\ne0$, which would preserve some of these objects in the parameterized case. 

We proceed with the computation of Lie symmetries for the case $J=0$ using the infinitesimal invariance criterion \cite{olve86Ay,ovsi82a}.
We look for the infinitesimal generators of one-parameter local symmetry groups of the primitive equations~\eqref{eq:PrimitiveEquations},
which constitute a Lie algebra~$\mathfrak g_f$ called the maximal Lie invariance algebra of these equations.
The subscript~$f$ indicates that in fact the algebra~$\mathfrak g_f$ depends on the the Coriolis parameter~$f$. 
It also depends on $(R,c_p)$ but this dependence is not significant for the further consideration. 
Each of the infinitesimal generators is a vector field of the form
\[
 Q=\tau\p_t + \xi^x\p_x+\xi^y\p_y+\xi^p\p_{p} +\eta^u\p_u+\eta^v\p_v+\xi^\omega\p_\omega+\eta^\phi\p_\phi+\eta^T\p_T,
\]
whose coefficients satisfy the system of determining equations implied by the infinitesimal invariance criterion.
We have computed the algebra~$\mathfrak g_f$ using the Maple package \texttt{DESOLV}~\cite{butc03Ay,carm00Ay,vu07Ay}.
Splitting the general expression for $Q$ with respect to algebra's parameters gives the following vector fields spanning the algebra~$\mathfrak g_f$:
\begin{align}\label{eq:MaximalLieInvarianceAlgebraPrimitiveEquationsF}
\begin{split}
 & \DD_1=t\p_t+\hat fty\p_x-\hat ftx\p_y-(u-\hat ftv-\hat fy)\p_u-(v+\hat ftu+\hat fx)\p_v-\omega\p_\omega{}\\
 &\phantom{\DD_1=}{}-\big(2\phi+\hat f^2(x^2+y^2)\big)\p_\phi-2T\p_T,\\
 &\DD_2=x\p_x+y\p_y+u\p_u+v\p_v+2\phi\p_\phi+2T\p_T,\quad
 \DD_3=p\p_p+\omega\p_\omega,\\
 &\PP=\p_t,\quad \JJ=-y\p_x+x\p_y-v\p_u+u\p_v,\quad \SSS=p^{\kappa}(c_p\p_\phi-\p_T),\\
 &\XX(\boldsymbol\gamma)=\boldsymbol\gamma\cdot\p_{\mathbf x}+\boldsymbol\gamma_t\cdot\p_{\mathbf v}
 -\big(\boldsymbol\gamma_{tt}\cdot\mathbf x+f(\gamma^1_ty-\gamma^2_tx)\big)\p_\phi,\quad
 \ZZ(\alpha)=\alpha\p_\phi,
\end{split}
\end{align}
where $\hat f:=f/2$, $\mathbf x:=(x,y)^{\mathsf T}$, $\p_{\mathbf x}:=(\p_x,\p_y)^{\mathsf T}$, 
$\p_{\mathbf v}:=(\p_u,\p_v)^{\mathsf T}$, $\boldsymbol\gamma:=(\gamma^1,\gamma^2)^{\mathsf T}$,
and the parameter functions $\gamma^1$, $\gamma^2$ and $\alpha$ run through the set of smooth functions depending on $t$.
The associated one-parameter groups consist of (i)--(iii) scalings, (iv) time translations, (v)~planar rotations, 
(vi) simultaneous shifts of the geopotential and temperature proportional to $p^\kappa$,
(vii) generalized shifts in the space variables, which generalize the space translations and Galilean boosts, 
(viii) gauging of the geopotential, which depend on~$t$.

The algebra $\mathfrak g_f$ is not singular in the Coriolis parameter~$f$ at $f=0$, which means that it is possible to set $f=0$ in~\eqref{eq:MaximalLieInvarianceAlgebraPrimitiveEquationsF}. The remaining question is whether there are additional infinitesimal generators extending the algebra $\mathfrak g_0$ when $f=0$ in~\eqref{eq:PrimitiveEquations}. Computing Lie symmetries of system~\eqref{eq:PrimitiveEquations} for $f=0$ shows that this is not the case, i.e., the primitive equations in a rest reference frame admit $\mathfrak g_0$ as the maximal Lie invariance algebra, which is spanned by the vector fields 
\begin{align}\label{eq:MaximalLieInvarianceAlgebraPrimitiveEquationsFzero}
\begin{split}
 &\DD_1=t\p_t- u\p_u- v\p_v-\omega\p_\omega-2\phi\p_\phi-2 T\p_T,\\
 &\DD_2= x\p_x+ y\p_y+ u\p_u+ v\p_v+2\phi\p_\phi+2 T\p_T,\quad
 \DD_3= p\p_p+\omega\p_\omega,\\
 &\PP=\p_t,\quad
 \JJ=-y\p_x+ x\p_y- v\p_u+ u\p_v,\quad
 \SSS=p^{\kappa}(c_p\p_\phi-\p_T),\\
 &\XX(\boldsymbol\gamma)=\boldsymbol\gamma\cdot\p_{\mathbf x}+\boldsymbol\gamma_t\cdot\p_{\mathbf v}-\boldsymbol\gamma_{tt}\cdot\mathbf x\p_\phi,\quad
 \ZZ(\alpha)=\alpha\p_\phi
\end{split}
\end{align}
with the same parameters as in~\eqref{eq:MaximalLieInvarianceAlgebraPrimitiveEquationsF}.
Up to anti-symmetry of the Lie bracket of vector fields, the nonzero commutation relations among the vector fields spanning $\mathfrak g_0$ are exhausted by
\begin{align*}
&[\DD_1,\PP]=-\PP,\quad
 [\DD_1,\SSS] = 2\SSS,\quad
 [\DD_1,\XX(\boldsymbol\gamma)]=\XX(t\boldsymbol\gamma_t),\quad
 [\DD_1,\ZZ(\alpha)]=\ZZ(t\alpha_t+2\alpha),\\
&[\DD_2,\SSS]=-2\SSS,\quad
 [\DD_2,\XX(\boldsymbol\gamma)]=- \XX(\boldsymbol\gamma),\quad
 [\DD_2,\ZZ(\alpha)]=-2\ZZ(\alpha),\quad
 [\DD_3,\SSS]=\kappa\SSS,\\
&[\PP,\XX(\boldsymbol\gamma)]=\XX(\boldsymbol\gamma_t),\quad
 [\PP,\ZZ(\alpha)]=\ZZ(\alpha_t),\quad
 [\JJ,\XX(\boldsymbol\gamma)]=\XX(\gamma^2,-\gamma^1),\\
&[\XX(\boldsymbol\gamma),\XX(\boldsymbol\sigma)]=\ZZ(\boldsymbol\sigma\cdot\boldsymbol\gamma_{tt}-\boldsymbol\gamma\cdot\boldsymbol\sigma_{tt}), 
\end{align*}
where $\boldsymbol\sigma:=(\sigma^1,\sigma^2)^{\mathsf T}$ is one more two-dimensional vector-valued smooth function of~$t$.
Based on the above commutation relations,
one can see that the Lie algebra $\mathfrak g_0$ has the structure 
$\mathfrak g_0=(\mathfrak s_2\oplus \mathfrak s_3)\lsemioplus \mathfrak i,$ 
where
$\mathfrak s_2=\langle\p_t,\DD_{1}\rangle$ is a realization of the two-dimensional nonabelian algebra,
$\mathfrak s_3=\langle\DD_{2},\DD_{3},\JJ\rangle$ is a  realization of the three-dimensional abelian algebra and
$\mathfrak i=\langle\XX(\boldsymbol\gamma),\ZZ(\alpha),\SSS\rangle$ is an infinite-dimensional ideal in~$\mathfrak g_0$,
$\mathfrak i=\big(\langle\XX(\boldsymbol\gamma)\rangle\lsemioplus\langle\ZZ(\alpha)\rangle\big)\oplus\langle\SSS\rangle$,
and $\langle\ZZ(\alpha)\rangle$ and $\langle\SSS\rangle$ are abelian ideals in the entire algebra~$\mathfrak g_0$.

Upon redefining the vector fields spanning the algebra $\mathfrak g_f$ according to
\begin{gather}\label{eq:RedefiningVFsOfGf}
\PP\rightsquigarrow\check\PP:=\p_t-\hat f\JJ,\quad \XX(\boldsymbol\gamma)\rightsquigarrow\check\XX(\boldsymbol\gamma):=\XX(\check{\boldsymbol\gamma}),
\end{gather}
where $\check{\boldsymbol\gamma}=(\gamma^1\cos(\hat ft)+\gamma^2\sin(\hat ft),-\gamma^1\sin(\hat ft)+\gamma^2\cos(\hat ft))$ and the remaining vector fields from~\eqref{eq:MaximalLieInvarianceAlgebraPrimitiveEquationsF} are left unchanged, they satisfy the same commutation relations as the vector fields~\eqref{eq:MaximalLieInvarianceAlgebraPrimitiveEquationsFzero}. Therefore, the algebras $\mathfrak g_f$ and $\mathfrak g_0$ are isomorphic, which is a necessary condition for the existence of a point transformation mapping the primitive equations with $f\ne0$ to the primitive equation in a resting reference frame ($f=0$)~\cite{blum89Ay}. This allows us to use the algebraic method for finding the transformation relating the two systems with $f=0$ and $f\ne0$ to each other.

Suppose that a point transformation
\begin{equation}\label{EqGenFormOfPointTransForPrimitiveEqs}
\mathcal T\colon\quad \tilde z^i=\mathcal T^i(t,x,y,p,u,v,\omega,\phi,T),\quad i\in\{t,x,y,p,u,v,\omega,\phi,T\},
\end{equation}
realizes the above isomorphism between the algebras $\mathfrak g_f$ and $\mathfrak g_0$, $\mathcal T_*\mathfrak g_f=\mathfrak g_0$. 
Here and in what follows
\begin{gather*}
(z^t,z^x,z^y,z^p,z^u,z^v,z^\omega,z^\phi,z^T)=(t,x,y,p,u,v,\omega,\phi,T),\\
(\tilde z^t,\tilde z^x,\tilde z^y,\tilde z^p,\tilde z^u,\tilde z^v,\tilde z^\omega,\tilde z^\phi,\tilde z^T)=
(\tilde t,\tilde x,\tilde y,\tilde p,\tilde u,\tilde v,\tilde \omega,\tilde \phi,\tilde T).
\end{gather*}
Then, the relation $\mathcal T_*Q=\tilde Q$ upon the corresponding vector fields $Q\in\mathfrak g_f$ and $\tilde Q\in\mathfrak g_0$
according to~\eqref{eq:RedefiningVFsOfGf} reads
\begin{equation}\label{eq:TransformationRuleVectorFields}
 Q\mathcal T^i = \mathcal T^*(\tilde Q\tilde z^i),\quad i\in\{t,x,y,p,u,v,\omega,\phi,T\},
\end{equation}
which is the standard rule for pushing forward vector fields by a point transformation.
Evaluating~\eqref{eq:TransformationRuleVectorFields} for $Q\in\big\{\ZZ(1),\ZZ(t)\big\}$, 
we derive \smash{$\mathcal T^i_\phi=0$}, $i\in\{x,y,p,u,v,\omega,T\}$, 
\smash{$\mathcal T^\phi_\phi=1$} and $\mathcal T^t=t$. 
Then the condition~\eqref{eq:TransformationRuleVectorFields} with $Q=\SSS$ implies that 
\smash{$\mathcal T^i_T=0$}, $i\in\{x,y,p,u,v,\omega\}$, 
$p^\kappa\mathcal T^T_T=(\mathcal T^p)^\kappa$ and 
$p^\kappa(c_p-\mathcal T^\phi_T)=c_p(\mathcal T^p)^\kappa$.
Since the parameter~$\boldsymbol\gamma$ is an arbitrary smooth vector-valued function of~$t$, 
we can split the condition~\eqref{eq:TransformationRuleVectorFields} with $Q=\check\XX(\boldsymbol\gamma)$ 
with respect to the components of~$\boldsymbol\gamma$ and their derivatives, 
obtaining 
\begin{gather*}
\mathcal T^p_j=\mathcal T^\omega_j=\mathcal T^T_j=0,\ j\in\{x,y,u,v\},\\ 
\mathcal T^x=\cos(\hat ft)x-\sin(\hat ft)y,\quad
\mathcal T^y=\sin(\hat ft)x+\cos(\hat ft)y,\\
 \mathcal T^u_u=\mathcal T^v_v=\cos(\hat ft),\quad 
-\mathcal T^u_v=\mathcal T^v_u=\sin(\hat ft),\\    
\mathcal T^u_x=\mathcal T^v_y=-\hat f\sin(\hat ft),\quad 
-\mathcal T^u_y=\mathcal T^v_x=\hat f\cos(\hat ft),\\ 
\mathcal T^\phi_x=\hat f^2x,\quad
\mathcal T^\phi_y=\hat f^2y,\quad 
\mathcal T^\phi_u=\mathcal T^\phi_v=0.
\end{gather*}
In view of the derived equations, the condition~\eqref{eq:TransformationRuleVectorFields} with $Q=\DD_2$ leads to
\begin{gather*}
\mathcal T^u=\cos(\hat ft)u-\sin(\hat ft)v-\hat f(\sin(\hat ft)x+\cos(\hat ft)y),\\
\mathcal T^v=\sin(\hat ft)u+\cos(\hat ft)v+\hat f(\cos(\hat ft)x-\sin(\hat ft)y),\\
2\mathcal T^\phi=2\phi+\hat f^2(x^2+y^2)+T\mathcal T^\phi_T,\quad
T\mathcal T^T_T=\mathcal T^T, 
\end{gather*}
whereas for $Q=\check P$ the condition~\eqref{eq:TransformationRuleVectorFields} gives
$\mathcal T^p_t=\mathcal T^\omega_t=\mathcal T^T_t=\mathcal T^\phi_t=0$. 
Successively considering $Q=\DD_1$ and $Q=\DD_3$, we obtain 
$\mathcal T^p_\omega=\mathcal T^\phi_\omega=\mathcal T^T_\omega=0$, $\omega\mathcal T^\omega_\omega=\mathcal T^\omega$ and 
$p\mathcal T^p_p=\mathcal T^p$, $\mathcal T^\omega_p=\mathcal T^\phi_p=\mathcal T^T_p=0$.
Therefore, $\mathcal T^p=c_1p$ and $\mathcal T^\omega=c_2\omega$ with nonzero constants~$c_1$ and~$c_2$.
Setting $c_1=c_2=1$ results in $\mathcal T^p=p$, $\mathcal T^\omega=\omega$, $\mathcal T^T=T$ and $\mathcal T^\phi_T=0$.
It can be checked that then the condition~\eqref{eq:TransformationRuleVectorFields} 
with the constructed transformation~$\mathcal T$ is identically satisfied by any $Q\in\mathfrak g_f$.
Moreover, the direct substitution shows that the same transformation also relates the primitive equations with $f\ne0$ to that with $f=0$. 
This proves the following theorem.

\begin{theorem}\label{thm:ReductionOfPrimitiveEquationsToReferenceFrameAtRest}
 The isentropic primitive equations~\eqref{eq:PrimitiveEquations} in a reference frame with constant rotation can be transformed to the isentropic primitive equations in a reference frame at rest (i.e., $f=0$) upon using the transformation
 \begin{align}\label{eq:PrincipalTransformationPrimitiveEquations}
 \begin{split}
  &\tilde t=t,\quad 
   \tilde x=\cos(\hat ft)x-\sin(\hat ft)y,\quad 
   \tilde y=\sin(\hat ft)x+\cos(\hat ft)y,\quad 
   \tilde p = p,\\
  &\tilde u=\cos(\hat ft)u-\sin(\hat ft)v-\hat f(\sin(\hat ft)x+\cos(\hat ft)y),\\ 
  &\tilde v=\sin(\hat ft)u+\cos(\hat ft)v+\hat f(\cos(\hat ft)x-\sin(\hat ft)y),\\
  &\tilde\omega=\omega,\quad 
   \tilde\phi =\phi+\frac{f^2}{8}(x^2+y^2),\quad 
   \tilde T=T,
 \end{split}
 \end{align}
  where $\hat f:=f/2$. The same transformation maps the maximal Lie invariance algebra $\mathfrak g_f$ to $\mathfrak g_0$.
\end{theorem}

In Theorem~\ref{thm:ReductionOfPrimitiveEquationsToReferenceFrameAtRest} 
and in Remark~\ref{rem:PrincipalTransformationPrimitiveEquationsInCylCoords} below,
we assume that the variables without tildes (resp.\ with tildes) are related to the system with $f\neq0$ (resp.\ $f=0$).

\begin{remark}\label{rem:PrincipalTransformationPrimitiveEquationsInCylCoords}
 In the cylindrical coordinates $(r,\theta,p)$, the transformation~\eqref{eq:PrincipalTransformationPrimitiveEquations} takes the particularly simple form
 \begin{align*}
   &\tilde t=t,\quad \tilde r = r,\quad \tilde\theta = \theta + \frac{f}{2}t,\quad \tilde p=p,\\
   &\tilde u^r = u^r,\quad \tilde u^\theta = u^\theta + \frac{f}{2}r,\quad \tilde\omega = \omega,\quad \tilde\phi = \phi+\frac{f^2}{8}r^2,\quad \tilde T=T,
 \end{align*}
 where $u^r$ and $u^\theta$ are the velocity components in the radial and in the azimuthal directions, respectively.
\end{remark}

\begin{remark}
In the case $c_p=R$, i.e., $\kappa=1$, the system of primitive equations~\eqref{eq:PrimitiveEquations}, 
where we set $f=0$ without loss of generality, admits a wider maximal Lie invariance algebra~$\hat{\mathfrak g}_0$ than~$\mathfrak g_0$. 
In comparison with~$\mathfrak g_0$, additional spanning vector fields in~$\hat{\mathfrak g}_0$ are
\begin{align*}
  \mathcal R(\lambda)={}&2\lambda\p_t+\lambda_tx\p_x+\lambda_ty\p_y-2\lambda_tp\p_p
  -(\lambda_tu-\lambda_{tt}x)\p_u-(\lambda_tv-\lambda_{tt}y)\p_v\\
  &{}-(4\lambda_t\omega+2\lambda_{tt}p)\p_\omega-
  \left(2\lambda_t\phi+\frac12\lambda_{ttt}(x^2+y^2)\right)\p_\phi
  -2\lambda_tT\p_T,\\
  \mathcal P(\psi)={}&\psi\p_p+\psi_t\p_\omega+\frac{\psi}pT\p_T,
\end{align*}
where $\lambda$ and $\psi$ run through the set of smooth functions of $t$. 
It is clear that the vector field $\mathcal R(\lambda)$ is a generalization 
of the usual shifts in $t$ ($\lambda=\const$) and the scaling vector field $2\DD_1+\DD_2$ ($\lambda=t$). 
For arbitrary $\lambda$, the transformations associated with the vector field~$\mathcal R(\lambda)$ can be interpreted as re-parameterization of time. 
The vector field~$\mathcal P(\psi)$ in turn corresponds to generalized Galilean boosts in the $p$-direction. 
At the same time, the case $c_p=R$ is unphysical as $c_p>R$.
\end{remark}

So as to derive the transformation~\eqref{eq:PrincipalTransformationPrimitiveEquations} using the algebraic method it was necessary to assume $J=0$, i.e.\ the system was required to be isentropic. 
This assumption is crucial as for general values of the parameter-function $J=J(t,x,y,p)$ 
the primitive equations~\eqref{eq:PrimitiveEquations} only admit the span of the gauging vector fields $\ZZ(\alpha)$ and $\mathcal S$ 
as their maximal Lie invariance algebra. 
This span is not enough to derive a sufficient number of equations for the components of the transformation~\eqref{eq:PrincipalTransformationPrimitiveEquations}. 
On the other hand, one can check the validity of this transformation for the case $J\ne0$ by direct computation. 
As the differential operator $\p_t+\vv\cdot\nabla$ is invariant with respect to the transformation~\eqref{eq:PrincipalTransformationPrimitiveEquations} 
and thus the right hand side of the temperature equation preserves its form, 
the same transformation also maps the primitive equations for $J\ne0$ in a rotating reference frame 
to the corresponding system in a resting reference frame with~$\tilde J$, where $\tilde J(\tilde t,\tilde x,\tilde y,\tilde p)=J(t,x,y,p)$.

\begin{corollary}
The transformation \eqref{eq:PrincipalTransformationPrimitiveEquations} maps 
the non-isentropic ($J\ne0$) primitive equations in a reference frame with constant rotation 
to the non-isentropic primitive equations in a reference frame at rest, 
where $\tilde J(\tilde t,\tilde x,\tilde y,\tilde p)=J(t,x,y,p)$.
\end{corollary}

\begin{remark}
 Owing to the invariance of the Lagrangian time derivative $\p_t+\vv\cdot\nabla$ 
under the transformation~\eqref{eq:PrincipalTransformationPrimitiveEquations}, 
it is possible to extend the system of primitive equations~\eqref{eq:PrimitiveEquations} 
by equations of the form
\[
S_t + \vv\cdot\nabla S + \omega S_p = Q
\]
without introducing new nontrivial transformation components 
for the prognostic variable $S$ and the source term $Q$, i.e., $\tilde S=S$ and $\tilde Q=Q$. 
Examples for physically relevant equations of the above form are, e.g., 
the moisture equation or any equation for a passively transported atmospheric tracer.
\end{remark}

\section{Complete point symmetry group}\label{sec:ompletePointSymmetryGroupPrimitiveEquations}

The above consideration shows that without loss of generality it suffices
to carry out group analysis of the primitive equations~\eqref{eq:PrimitiveEquations} only for the case $f=0$.
In this section we find the complete point symmetry group~$G_0$ of Eqs.~\eqref{eq:PrimitiveEquations} 
with $f=0$ and $J=0$ by the algebraic method proposed in~\cite{bihl11Cy}.
This method may be treated as an enhancement of the approach suggested in~\cite{hydo00By,hydo00Ay}
(see also~\cite{gray13Ay}) by embedding the notion of megaideals~\cite{popo03Ay},
which is a brief name for fully characteristic ideals. 
The enhanced method was further advanced in~\cite{bihl2015a,card12Ay,malt2021a}.
Its main benefit is that it can be applied even to
systems of differential equations possessing infinite-dimensional Lie invariance algebras,
which is the case for Eqs.~\eqref{eq:PrimitiveEquations} with $J=0$,
although it can also simplify analogous computations for Lie invariance algebras of finite but high dimensions.

The algebra~$\mathfrak g=\mathfrak g_0$ has the following obvious megaideals:
\begin{gather*}
\mathfrak g'=\langle\PP,\SSS,\XX(\boldsymbol\gamma),\ZZ(\alpha)\rangle,\quad
\mathfrak g''=\langle\XX(\boldsymbol\gamma),\ZZ(\alpha)\rangle,\quad
\mathfrak g'''=\mathrm Z_{\mathfrak g''}=\langle\ZZ(\alpha)\rangle,
\\
\mathrm Z_{\mathfrak g'}=\langle\SSS,\ZZ(1)\rangle,\quad
\mathrm Z_{\mathfrak g'}\cap\mathfrak g'''=\langle\ZZ(1)\rangle,\\
\mathfrak m_1=\mathrm C_{\mathfrak g}(\mathfrak g'')=\langle\DD_3,\SSS,\ZZ(\alpha)\rangle,\quad
\mathfrak m_1'=\langle\SSS\rangle,\quad
\mathrm C_{\mathfrak g}(\mathfrak m_1)=\langle\JJ,\XX(\boldsymbol\gamma),\ZZ(\alpha)\rangle,
\end{gather*}
where $\mathfrak a'$, $\mathrm Z_{\mathfrak a}$ and $\mathrm C_{\mathfrak a}(\mathfrak b)$
denote the derivative and the center of a Lie algebra~$\mathfrak a$
and the centralizer of a subalgebra~$\mathfrak b$ in~$\mathfrak a$, respectively.
Here and in what follows the parameters $\gamma^1$, $\gamma^2$ and $\alpha$ run through the set of smooth functions depending on $t$.

To find more megaideals of~$\mathfrak g$, we apply Proposition~1 from~\cite{card12Ay}
for various special choices of the megaideals~$\mathfrak i_0$, $\mathfrak i_1$ and $\mathfrak i_2$ of~$\mathfrak g$.
This proposition states that the set~$\mathfrak s$ of elements from~$\mathfrak i_0$
whose commutators with arbitrary elements from~$\mathfrak i_1$ belong to~$\mathfrak i_2$
is also a megaideal of~$\mathfrak g$.
Thus, for $\mathfrak i_0=\mathfrak g'''$, $\mathfrak i_1=\mathfrak g'$ and $\mathfrak i_2=\mathrm Z_{\mathfrak g'}\cap\mathfrak g'''=\langle\ZZ(1)\rangle$,
we obtain $\mathfrak s=\langle\ZZ(1),\ZZ(t)\rangle$ and hence this is a megaideal.
We reassign the last~$\mathfrak s$ as $\mathfrak i_2$ and iterate the procedure with the same~$\mathfrak i_0$ and~$\mathfrak i_1$,
which gives the series of megaideals $\langle\ZZ(1),\ZZ(t),\dots,\ZZ(t^n)\rangle$, $n\in\mathbb N_0$.

A convenient choice for~$\mathfrak i_0$ and $\mathfrak i_1$ is $\mathfrak i_0=\mathfrak i_1=\mathfrak g$
when $\mathfrak i_2$ is varying.
For $\mathfrak i_2=\mathfrak m_1'$ and $\mathfrak i_2=\mathfrak g''$
we respectively have the megaideals
\[\mathfrak s=\langle\DD_3,\SSS\rangle=:\mathfrak m_2 
\ \ \mbox{and}\ \ 
\mathfrak s=\langle\kappa\DD_2+2\DD_3,\JJ,\XX(\boldsymbol\gamma),\ZZ(\alpha)\rangle.\]
Then 
\[
\mathrm C_{\mathfrak g'}(\mathfrak m_2)=\big\langle\PP,\XX(\boldsymbol\gamma),\ZZ(\alpha)\big\rangle
\ \ \mbox{and}\ \ 
\mathrm C_{\mathfrak g}(\mathfrak m_2)=\big\langle\DD_1+\DD_2,\DD_2+2\DD_3,\JJ,\PP,\XX(\boldsymbol\gamma),\ZZ(\alpha)\big\rangle=:\mathfrak m_3
\]
are also megaideals, as well as
\[\mathrm C_{\mathfrak m_3}(\mathrm Z_{\mathfrak g'}\cap\mathfrak g''')=\big\langle\DD_1+\DD_2,\JJ,\PP,\XX(\boldsymbol\gamma),\ZZ(\alpha)\big\rangle.\]

Applying again Proposition~1 from~\cite{card12Ay} on the next step,
we take $\mathfrak i_0=\mathfrak i_1=\mathrm C_{\mathfrak g'}(\mathfrak m_2)$ and $\mathfrak i_2=\mathfrak g'''$
and derive the megaideal $\mathfrak s=\langle\XX(1,0),\XX(0,1),\ZZ(\alpha)\rangle=:\mathfrak m_4$.
We reassign the last~$\mathfrak s$ as $\mathfrak i_2$ and iterate the procedure with the same~$\mathfrak i_0$ and~$\mathfrak i_1$,
which gives the series of megaideals
\[\big\langle\XX(1,0),\XX(0,1),\XX(t,0),\XX(0,t),\dots,\XX(t^n,0),\XX(0,t^n),\ZZ(\alpha)\big\rangle,\quad n\in\mathbb N_0.\]
Considering $\mathfrak i_0=\mathfrak g$ and $\mathfrak i_1=\mathfrak m_4\oplus\mathfrak m_1'$ with $\mathfrak i_2=\mathfrak g'''$,
we get $\mathfrak s=\big\langle\DD_1,\PP,\SSS,\XX(\boldsymbol\gamma),\ZZ(\alpha)\big\rangle$.

Some of the above megaideals of~$\mathfrak g_0$ can be neglected in the course of computing the complete point symmetry group~$G_0$
of the primitive equations~\eqref{eq:PrimitiveEquations} with $f=0$ by the algebraic method.
Indeed, the condition $G_*\mathfrak i\subseteq\mathfrak i$ for a megaideal~$\mathfrak i$
may only result in constraints for components of point symmetry transformations
that are consequences of those obtained in the course of the computation with other megaideals.
In particular, this is the case if a megaideal~$\mathfrak i$ is a sum of other megaideals.
To optimize the computation, we select a minimal set of megaideals
that allow us to easily derive a set of constraints for components of point symmetry transformations that is maximal within the algebraic framework.
We choose the following megaideals from those we have computed:
\begin{gather}\label{EqMegaIdealsForComputationOfPointSymGroupOfPrimitiveEqs}
\begin{split}&
\big\langle\ZZ(1)\big\rangle,\quad
\big\langle\ZZ(1),\ZZ(t)\big\rangle,\quad
\big\langle\SSS\big\rangle,\quad
\big\langle\XX(1,0),\XX(0,1),\ZZ(\alpha)\big\rangle,
\\[.5ex]&
\big\langle\XX(t,0),\XX(0,t),\XX(1,0),\XX(0,1),\ZZ(\alpha)\big\rangle,
\\[.5ex]&
\big\langle\XX(t^2,0),\XX(0,t^2),\XX(t,0),\XX(0,t),\XX(1,0),\XX(0,1),\ZZ(\alpha)\big\rangle,
\\[.5ex]&
\big\langle\JJ,\XX(\boldsymbol\gamma),\ZZ(\alpha)\big\rangle,\quad
\big\langle\PP,\XX(\boldsymbol\gamma),\ZZ(\alpha)\big\rangle,\quad
\big\langle\DD_1+\DD_2,\JJ,\PP,\XX(\boldsymbol\gamma),\ZZ(\alpha)\big\rangle,
\\[.5ex]&
\big\langle\DD_3,\SSS\big\rangle,\quad
\big\langle\DD_1,\PP,\SSS,\XX(\boldsymbol\gamma),\ZZ(\alpha)\big\rangle.
\end{split}
\end{gather}

The general form of point transformations
that acts in the space of the independent and dependent variables of the primitive equations~\eqref{eq:PrimitiveEquations}
is given by Eq.~\eqref{EqGenFormOfPointTransForPrimitiveEqs},
where the corresponding Jacobian~$\mathrm J$ does not vanish.
For a point transformation~$\mathcal T$ to be qualified as a point symmetry of the primitive equations~\eqref{eq:PrimitiveEquations} with~$f=0$,
its counterpart~$\mathcal T_*$ push-forwarding vector fields should preserve
each of the selected  megaideals~\eqref{EqMegaIdealsForComputationOfPointSymGroupOfPrimitiveEqs} of the algebra~$\mathfrak g_0$.
We have additionally ordered the megaideal list~\eqref{EqMegaIdealsForComputationOfPointSymGroupOfPrimitiveEqs} 
in such a way that megaideals heading the list give
more elementary equations of the form $\mathcal T^j_{z^i}=0$ with some $i,j\in\{t,x,y,p,u,v,\omega,\phi,T\}$
or allow us to specify the expressions for some $\mathcal T^j$.
As a result, we obtain the conditions
\begin{subequations}\label{EqMegaidealConstraintForPointSymTransOfPrimitiveEqs}
\begin{align}
&\mathcal T_* \ZZ(1)=\mathcal T^i_\phi\p_{\tilde z^i}=a_1\tilde\ZZ(1),
\label{EqMegaidealConstraintForPointSymTransOfPrimitiveEqsZ1}\\[.5ex]
&\mathcal T_* \ZZ(t)=t\mathcal T^i_\phi\p_{\tilde z^i}=a_2\tilde\ZZ(\tilde t)+a_3\tilde\ZZ(1),
\label{EqEqMegaidealConstraintForPointSymTransOfPrimitiveEqsZt}\\[.5ex]
&\mathcal T_* \SSS=p^\kappa(c_p\mathcal T^i_\phi-\mathcal T^i_T)\p_{\tilde z^i}=a_4\tilde\SSS,
\label{EqEqMegaidealConstraintForPointSymTransOfPrimitiveEqsS}\\[.5ex]
&\mathcal T_* \mathcal X(1,0)=\mathcal T^i_x\p_{\tilde z^i}=\tilde\XX(b^{00}_{11},b^{00}_{21})+\tilde\ZZ(\tilde\alpha^{01}),
\label{EqMegaidealConstraintForPointSymTransOfPrimitiveEqsX10}\\[.5ex]
&\mathcal T_* \mathcal X(0,1)=\mathcal T^i_y\p_{\tilde z^i}=\tilde\XX(b^{00}_{12},b^{00}_{22})+\tilde\ZZ(\tilde\alpha^{02}),
\label{EqMegaidealConstraintForPointSymTransOfPrimitiveEqsX01}\\[.5ex]
&\mathcal T_* \mathcal X(t,0)=(t\mathcal T^i_x+\mathcal T^i_u)\p_{\tilde z^i}=\tilde\XX(b^{11}_{11}\tilde t+b^{10}_{11},b^{11}_{21}\tilde t+b^{10}_{21})+\tilde\ZZ(\tilde\alpha^{11}),
\label{EqMegaidealConstraintForPointSymTransOfPrimitiveEqsXt0}\\[.5ex]
&\mathcal T_* \mathcal X(0,t)=(t\mathcal T^i_y+\mathcal T^i_v)\p_{\tilde z^i}=\tilde\XX(b^{11}_{12}\tilde t+b^{10}_{12},b^{11}_{22}\tilde t+b^{10}_{22})+\tilde\ZZ(\tilde\alpha^{12}),
\label{EqMegaidealConstraintForPointSymTransOfPrimitiveEqsX0t}\\[.5ex]
&\mathcal T_* \mathcal X(t^2,0)=(t^2\mathcal T^i_x+2t\mathcal T^i_u-2x\mathcal T^i_\phi)\p_{\tilde z^i}
\nonumber\\ &\phantom{\mathcal T_* \mathcal X(t^2,0)}
=\tilde\XX(b^{22}_{11}\tilde t^2+b^{21}_{11}\tilde t+b^{20}_{11},b^{22}_{21}\tilde t^2+b^{21}_{21}\tilde t+b^{20}_{21})+\tilde\ZZ(\tilde\alpha^{21}),
\label{EqMegaidealConstraintForPointSymTransOfPrimitiveEqsXt20}\\[.5ex]
&\mathcal T_* \mathcal X(0,t^2)=(t^2\mathcal T^i_y+2t\mathcal T^i_v-2y\mathcal T^i_\phi)\p_{\tilde z^i}
\nonumber\\ &\phantom{\mathcal T_* \mathcal X(0,t^2)}
=\tilde\XX(b^{22}_{12}\tilde t^2+b^{21}_{12}\tilde t+b^{20}_{12},b^{22}_{22}\tilde t^2+b^{21}_{22}\tilde t+b^{20}_{22})+\tilde\ZZ(\tilde\alpha^{22}),
\label{EqMegaidealConstraintForPointSymTransOfPrimitiveEqsX0t2}\\[.5ex]
&\mathcal T_* \JJ=(x\mathcal T^i_y-y\mathcal T^i_x+u\mathcal T^i_v-v\mathcal T^i_u)\p_{\tilde z^i}=a_5\tilde\JJ+\XX(\tilde{\boldsymbol\gamma}^3)+\tilde\ZZ(\tilde\alpha^3),
\label{EqMegaidealConstraintForPointSymTransOfPrimitiveEqsJ}\\[.5ex]
&\mathcal T_* \PP=\mathcal T^i_t\p_{\tilde z^i}=a_6\tilde\PP+\XX(\tilde{\boldsymbol\gamma}^4)+\tilde\ZZ(\tilde\alpha^4),
\label{EqMegaidealConstraintForPointSymTransOfPrimitiveEqsP}\\[.5ex]
&\mathcal T_* (\DD_1\!+\!\DD_2)=(t\mathcal T^i_t+x\mathcal T^i_x+y\mathcal T^i_y+\omega\mathcal T^i_\omega)\p_{\tilde z^i} =a_7(\tilde\DD_1\!+\!\tilde\DD_2)+a_8\tilde\PP+\XX(\tilde{\boldsymbol\gamma}^5)+\tilde\ZZ(\tilde\alpha^5),
\label{EqMegaidealConstraintForPointSymTransOfPrimitiveEqsD1D2}\\[.5ex]
&\mathcal T_* \DD_3=(p\mathcal T^i_p+\omega\mathcal T^i_\omega)\p_{\tilde z^i}=a_9\tilde\DD_3+a_{10}\tilde\SSS,
\label{EqMegaidealConstraintForPointSymTransOfPrimitiveEqsD3}\\[.5ex]
&\mathcal T_* \DD_1=(t\mathcal T^i_t-u\mathcal T^i_u-v\mathcal T^i_v-\omega\mathcal T^i_\omega-2\phi\mathcal T^i_\phi-2T\mathcal T^i_T)\p_{\tilde z^i}
\nonumber\\ &\phantom{\mathcal T_* \DD_1}
=a_{11}\tilde\DD_1+a_{12}\tilde\PP+a_{13}\tilde\SSS+\XX(\tilde{\boldsymbol\gamma}^6)+\tilde\ZZ(\tilde\alpha^6),
\label{EqMegaidealConstraintForPointSymTransOfPrimitiveEqsD1}
\end{align}
\end{subequations}
where $i\in\{t,x,y,p,u,v,\omega,\phi,T\}$, and we assume summation with respect to repeated indices;
$a_s$, $s=1,\dots,13$, $b^{00}_{kl}$, $b^{10}_{kl}$, $b^{11}_{kl}$, $b^{20}_{kl}$, $b^{21}_{kl}$ and $b^{22}_{kl}$, $k,l=1,2$, are constants;
$\tilde{\boldsymbol\gamma}^m=(\tilde\gamma^{m1},\tilde\gamma^{m2})^{\mathsf T}$, $m=3,\dots,6$,
and the parameters $\tilde\alpha^{0l}$, $\tilde\alpha^{1l}$, $\tilde\alpha^{2l}$,
$\tilde\gamma^{ml}$, $\tilde\gamma^{ml}$ and $\tilde\alpha^m$ are smooth functions depending on~$\tilde t$.

We will derive constraints on~$\mathcal T$ by sequentially equating the coefficients of vector fields
in the conditions \eqref{EqMegaidealConstraintForPointSymTransOfPrimitiveEqs}
and by taking into account the constraints obtained in previous steps.

Thus, the condition~\eqref{EqMegaidealConstraintForPointSymTransOfPrimitiveEqsZ1} directly implies that
$\mathcal T^\phi_\phi=a_1$ and $\mathcal T^i_\phi=0$ if $i\ne\phi$.
Then the constant~$a_1$ is nonzero since the Jacobian~$\mathrm J$ does not vanish.
The equation $\smash{a_1t=a_2\tilde t+a_3}$ derived from the condition~\eqref{EqEqMegaidealConstraintForPointSymTransOfPrimitiveEqsZt}
gives that $a_2\ne0$ and hence the component~$\mathcal T^t$ depends only on~$t$ and the dependence is affine,
\[\mathcal T^t=a_1a_2^{-1}t-a_3a_2^{-1}.\]
This completely specifies expression for~$\mathcal T^t$ and also implies that $\p_{\tilde t}=a_1^{-1}a_2\p_t$.

The condition~\eqref{EqEqMegaidealConstraintForPointSymTransOfPrimitiveEqsS} is split into the equations
\[
p^\kappa(c_p\mathcal T^T_\phi-\mathcal T^T_T)=-a_4(\mathcal T^p)^\kappa,\quad
p^\kappa(c_p\mathcal T^\phi_\phi-\mathcal T^\phi_T)=a_4c_p(\mathcal T^p)^\kappa,\quad 
c_p\mathcal T^i_\phi-\mathcal T^i_T=0, \ i\ne\phi,T.
\]
Therefore,
$\mathcal T^T_T=a_4(\mathcal T^p/p)^\kappa$,
$\mathcal T^\phi_T=c_pa_1-c_pa_4(\mathcal T^p/p)^\kappa$,
$\mathcal T^i_T=0$ for $i\ne\phi,T$, and $a_4\ne0$.

Considering simultaneously the pairs of the conditions
\eqref{EqMegaidealConstraintForPointSymTransOfPrimitiveEqsX10} and~\eqref{EqMegaidealConstraintForPointSymTransOfPrimitiveEqsX01},
\eqref{EqMegaidealConstraintForPointSymTransOfPrimitiveEqsXt0} and~\eqref{EqMegaidealConstraintForPointSymTransOfPrimitiveEqsX0t}, as well as
\eqref{EqMegaidealConstraintForPointSymTransOfPrimitiveEqsXt20} and~\eqref{EqMegaidealConstraintForPointSymTransOfPrimitiveEqsX0t2},
we derive that
\begin{gather*}\arraycolsep=0ex
\begin{array}{l}
\mathcal T^x_x=b^{00}_{11},\ \mathcal T^y_x=b^{00}_{21},\quad \mathcal T^\phi_x=\tilde\alpha^{01},\\[1ex]
\mathcal T^x_y=b^{00}_{12},\ \mathcal T^y_y=b^{00}_{22},\quad \mathcal T^\phi_y=\tilde\alpha^{02},
\end{array}\quad
\mathcal T^i_x=\mathcal T^i_y=0, \ i\ne x,y,\phi;
\\[1ex]
\mathcal T^x_u=b^{11}_{11}\tilde t+b^{10}_{11}-b^{00}_{11}t,\ \mathcal T^y_u=b^{11}_{21}\tilde t+b^{10}_{21}-b^{00}_{21}t,\quad
\mathcal T^u_u=b^{11}_{11},\ \mathcal T^v_u=b^{11}_{21},\quad \mathcal T^\phi_u=\tilde\alpha^{11}-t\tilde\alpha^{01},\\
\mathcal T^x_v=b^{11}_{12}\tilde t+b^{10}_{12}-b^{00}_{12}t,\ \mathcal T^y_v=b^{11}_{22}\tilde t+b^{10}_{22}-b^{00}_{22}t,\quad
\mathcal T^u_v=b^{11}_{12},\ \mathcal T^v_v=b^{11}_{22},\quad \mathcal T^\phi_v=\tilde\alpha^{12}-t\tilde\alpha^{02},\\
\mathcal T^i_u=\mathcal T^i_v=0,\ i=t,p,\omega,T;
\\[1ex]
b^{00}_{kl}t^2+2t(b^{11}_{kl}\tilde t+b^{10}_{kl}-b^{00}_{kl}t)=b^{22}_{kl}\tilde t^2+b^{21}_{kl}\tilde t+b^{20}_{kl},\quad
2b^{11}_{kl}t=2b^{22}_{kl}\tilde t+b^{21}_{kl},\quad k,l=1,2,
\\
2t\tilde\alpha^{11}-t^2\tilde\alpha^{01}-2a_1x=-2b^{22}_{11}\tilde x-2b^{22}_{21}\tilde y+\tilde\alpha^{21},\\
2t\tilde\alpha^{12}-t^2\tilde\alpha^{02}-2a_1y=-2b^{22}_{12}\tilde x-2b^{22}_{22}\tilde y+\tilde\alpha^{22}.
\end{gather*}
The last two equations imply that $|b^{22}_{kl}|_{k,l=1,2}^{}\ne0$ (otherwise, the Jacobian~$\mathrm J$ equals zero)
and thus the transformation components~$\tilde x=\mathcal T^x$ and~$\tilde y=\mathcal T^y$ depend only on~$(t,x,y)$.
More precisely, in terms of the constants~$b^{00}_{kl}$ we have the representation
\[
\mathcal T^x=b^{00}_{11}x+b^{00}_{12}y+\beta^1(t), \quad
\mathcal T^y=b^{00}_{21}x+b^{00}_{22}y+\beta^2(t),
\]
where $\beta^k$ are smooth functions of~$t$.
As $\mathcal T^x_u=\mathcal T^y_u=\mathcal T^x_v=\mathcal T^y_v=0$,
we obtain $b^{11}_{kl}\tilde t+b^{10}_{kl}-b^{00}_{kl}t=0$.
Then $b^{00}_{kl}=a_1a_2^{-1}b^{11}_{kl}$ and $b^{00}_{kl}=a_1^{\,2}a_2^{-2}b^{22}_{kl}$,
i.e., $B^{00}=a_1a_2^{-1}B^{11}$ and $B^{00}=a_1^{\,2}a_2^{-2}B^{22}$,
where we use the matrix notation $B^{00}=(b^{00}_{kl})$, $B^{11}=(b^{11}_{kl})$ and $B^{22}=(b^{22}_{kl})$.
On the other hand, $-2(B^{22})^{\mathrm T}B^{00}=-2a_1E$, where $E$ is the $2\times2$ unit matrix,
i.e., $(B^{00})^{\mathrm T}B^{00}=a_1^{\,3}a_2^{-2}E$,
which implies, e.g., for the $(1,1)$-entry that $(b^{00}_{11})^2+(b^{00}_{12})^2=a_1^{\,3}a_2^{-2}$.
Therefore, $a_1>0$ and thus we can represent the matrix~$B^{00}$ in the form
\[
B^{00}=a_1^{3/2}a_2^{-1}O,
\]
where $O$ is a $2\times2$ orthogonal matrix.
This completes specifying the expressions for~$\mathcal T^x$ and~$\mathcal T^y$.

The representation for~$B^{00}$ implies $b^{00}_{11}=b^{00}_{22}$ and $b^{00}_{12}=-b^{00}_{21}$.
Using this, we derive from the condition~\eqref{EqMegaidealConstraintForPointSymTransOfPrimitiveEqsJ} that
$B^{00}\mathbf x=a_5\tilde{\mathbf x}+(\tilde\gamma^{32},-\tilde\gamma^{31})^{\mathsf T}$,
which gives $a_5=1$, $\beta^1(t)=\tilde\gamma^{32}(\tilde t)$ and $\beta^2(t)=-\tilde\gamma^{31}(\tilde t)$.
In view of the above equations for derivatives $\mathcal T^j_{z^i}$ with $i,j=u,v$,
briefly representable as $(\mathcal T^j_{z^i})^{}_{i,j=u,v}=B^{11}$,
we also get from the condition~\eqref{EqMegaidealConstraintForPointSymTransOfPrimitiveEqsJ} that
\[a_1^{-1}a_2B^{00}\mathbf v=a_5\tilde{\mathbf v}+(\tilde\gamma^{32}_{\tilde t},-\tilde\gamma^{31}_{\tilde t})^{\mathsf T}.\]
Arranging the last equation results in finally specifying the expressions for~$\mathcal T^u$ and~$\mathcal T^v$,
\[
\mathcal T^u=\frac{a_2}{a_1}(b^{00}_{11}u+b^{00}_{12}v+\beta^1_t(t)), \quad
\mathcal T^v=\frac{a_2}{a_1}(b^{00}_{21}u+b^{00}_{22}v+\beta^2_t(t)).
\]
As the derivatives $\mathcal T^\phi_u$ and $\mathcal T^\phi_v$ may depend only on~$t$,
the condition~\eqref{EqMegaidealConstraintForPointSymTransOfPrimitiveEqsJ} gives that $\mathcal T^\phi_u=\mathcal T^\phi_v=0$.
The variables~$x$ and~$y$ are involved in the expression of~$\mathcal T^\phi$ only
within the summand $-a_2^{\,2}a_1^{-2}\boldsymbol\beta_{tt}\cdot B^{11}\mathbf x$, 
where $\boldsymbol\beta:=(\beta^1,\beta^2)^{\mathsf T}$.

The condition~\eqref{EqMegaidealConstraintForPointSymTransOfPrimitiveEqsP} obviously implies
the elementary equations $\mathcal T^p_t=\mathcal T^\omega_t=\mathcal T^T_t=0$
and the constraint that the transformation component~$\mathcal T^\phi$ may involve the variable~$t$
only via the above summand and one more summand that depends only on~$t$.

Two more elementary equations, $\mathcal T^p_\omega=\mathcal T^T_\omega=0$, follows from
the condition~\eqref{EqMegaidealConstraintForPointSymTransOfPrimitiveEqsD1D2}.
The equation implied by~\eqref{EqMegaidealConstraintForPointSymTransOfPrimitiveEqsD1D2} for~$\mathcal T^t$ is $t\mathcal T^t_t=a_7\mathcal T^t+a_8$,
which gives $a_7=1$.
Then the equation implied for~$\mathcal T^\omega$ takes he form $\omega\mathcal T^\omega_\omega=\mathcal T^\omega$.
The main feature of~$\mathcal T^\phi$ obtained from~\eqref{EqMegaidealConstraintForPointSymTransOfPrimitiveEqsD1D2}
is that the term $\omega\mathcal T^\phi_\omega$ depends only on~$t$, $x$ and~$y$.

Consider equations yielded by the condition~\eqref{EqMegaidealConstraintForPointSymTransOfPrimitiveEqsD3}.
Thus, the equation for~$\mathcal T^T$ is $p\mathcal T^T_p=-a_{10}p^\kappa$, and hence $\mathcal T^T_{pT}=0$.
As $\mathcal T^T_T=a_4(\mathcal T^p/p)^\kappa$, we get that $(\mathcal T^p/p)_p=0$, i.e. $p\mathcal T^p_p=\mathcal T^p$.
The equation for~$\mathcal T^p$ is $p\mathcal T^p_p=a_9\mathcal T^p$. Therefore, $a_9=1$.
Then the equation for~$\mathcal T^\omega$ takes the form $p\mathcal T^\omega_p+\omega\mathcal T^\omega_\omega=\mathcal T^\omega$
and, after combining with the analogous equation that is obtained from the condition~\eqref{EqMegaidealConstraintForPointSymTransOfPrimitiveEqsD1D2},
reduces to the equation $\mathcal T^\omega_p=0$.

Collecting coefficients of~$\p_{\tilde t}$, $\p_{\tilde T}$ and~$\p_{\tilde\phi}$ in~\eqref{EqMegaidealConstraintForPointSymTransOfPrimitiveEqsD1}
results in the equations $t\mathcal T^t_t=a_{11}\mathcal T^t+a_{12}$, $-2T\mathcal T^T_T=-2a_{11}\mathcal T^T-a_{13}(\mathcal T^p)^\kappa$ and
\[
t\mathcal T^\phi_t-\omega\mathcal T^\phi_\omega-\phi\mathcal T^\phi_\phi-2T\mathcal T^\phi_T=
-2a_{11}\mathcal T^\phi+a_{13}c_p(\mathcal T^p)^\kappa -\tilde{\boldsymbol\gamma}^6_{tt}\cdot\tilde{\mathbf x}+\tilde\alpha^6.
\]
The essential consequence of the first equation is $a_{11}=1$.
Then the third equation implies that $\mathcal T^\phi$ does not depend on and~$\omega$
since we have already proved that all summands in the left hand side of the equation as well as~$\mathcal T^p$ have this property.
From the second and third equations it is obvious that the variable~$p$ is involved in the expressions of~$\mathcal T^T$ and~$\mathcal T^\phi$ only
within the summands $-\frac12a_{13}(\mathcal T^p)^\kappa$ and $\frac12a_{13}c_p(\mathcal T^p)^\kappa$, respectively.

We introduce notation of the following constants:
\begin{gather*}
\varepsilon_0=-\frac{a_3}{a_2},\quad
\varepsilon_1= \frac{a_1}{a_2}\ne0,\quad
\varepsilon_2= \frac{a_1^{3/2}}{a_2}>0,\quad
\varepsilon_3= \frac{\mathcal T^p}p>0,\quad
\varepsilon_4= \frac12a_{12}\varepsilon_3^\kappa,\\
\varepsilon_5= \frac{\mathcal T^\omega}\omega\ne0,\quad
\varepsilon_6= a_{4}\varepsilon_3^\kappa.
\end{gather*}
The constant~$\varepsilon_3$ should be greater than zero for both physical and mathematical reasons
since the exponent $\varepsilon_3^\kappa$ should be well defined for all $\kappa\in(0,1)$
and, in view of the physical interpretation of the variable~$p$,
both its initial and transformed values should simultaneously be positive.
The constant~$\varepsilon_2$ can be assumed positive
since the parameters~$\varepsilon_2$ and~$O$ are defined up to simultaneously alternating their signs.
 Collecting all the restrictions we have derived for the components of the transformation~$\mathcal T$
within the algebraic approach and using the above notation,
we obtain the preliminary representation of this transformation,
\begin{gather}\label{eq:PreliminaryFormOfPointSymTransOfPrimitiveEqs}
\begin{split}
&\tilde t=\varepsilon_1t+\varepsilon_0, \quad
 \tilde{\mathbf x}=\varepsilon_2O\mathbf x+\boldsymbol\beta(t), \quad
 \tilde p=\varepsilon_3p, \\
&\tilde{\mathbf v}=\frac{\varepsilon_2}{\varepsilon_1}O\mathbf v+\frac1{\varepsilon_1}\boldsymbol\beta_t(t), \quad
 \tilde\omega=\varepsilon_5\omega, \\
&\tilde\phi=\frac{\varepsilon_2^2}{\varepsilon_1^2}\phi
  +c_p\left(\frac{\varepsilon_2^2}{\varepsilon_1^2}-\varepsilon_6\right)T+\varepsilon_4c_pp^\kappa
  -\frac{\varepsilon_2}{\varepsilon_1^2}\boldsymbol\beta_{tt}(t)\cdot O\mathbf x+\alpha(t), \quad
 \tilde T=\varepsilon_6T-\varepsilon_4p^\kappa.
\end{split}
\end{gather}

Not all parameters in the representation~\eqref{eq:PreliminaryFormOfPointSymTransOfPrimitiveEqs} are independent.
For the transformation~$\mathcal T$ to really be a point symmetry of the primitive equations~\eqref{eq:PrimitiveEquations},
some parameters have to satisfy additional constraints that cannot be derived within the framework of the algebraic approach.
This is why the computation should be completed by the direct method.
The application of the direct method can be simplified by factoring out \emph{a priori} known continuous transformations.
Thus, we can set $\varepsilon_0=\varepsilon_4=0$, $\boldsymbol\beta=\boldsymbol0$, $\alpha=0$
and $O$ to be equal to the diagonal matrix with the diagonal entries equal to~$-1$ or~$1$.

We calculate expressions for transformed derivatives and substitute them to the primitive equations~\eqref{eq:PrimitiveEquations}
written in terms of the transformed variables, which are with tildes.
Then we choose $u_t$, $v_t$, $\phi_p$, $\omega_p$ and~$T_t$ as principal derivatives,
express them in terms of other (parametric) derivatives from~\eqref{eq:PrimitiveEquations},
substitute the obtained expressions into the system derived on the previous step.
Splitting the resulting system with respect to parametric derivatives gives the missing equations,
\[
\varepsilon_5=\frac{\varepsilon_3}{\varepsilon_1},\quad
\varepsilon_6=\frac{\varepsilon_2^2}{\varepsilon_1^2}.
\]
This equations jointly with the representation~\eqref{eq:PreliminaryFormOfPointSymTransOfPrimitiveEqs} leads to the following assertion:

\begin{theorem}
The complete point symmetry (pseudo)group~$G_0$ of the primitive equations~\eqref{eq:PrimitiveEquations} 
with $f=0$, $J=0$ and arbitrary values $R>0$ and $\kappa\in(0,1)$ consists of the transformations
\begin{gather*}
\begin{split}
&\tilde t=\varepsilon_1t+\varepsilon_0, \quad
 \tilde{\mathbf x}=\varepsilon_2O\mathbf x+\boldsymbol\beta(t), \quad
 \tilde p=\varepsilon_3p, \\
&\tilde{\mathbf v}=\frac{\varepsilon_2}{\varepsilon_1}O\mathbf v+\frac1{\varepsilon_1}\boldsymbol\beta_t(t), \quad
 \tilde\omega=\frac{\varepsilon_3}{\varepsilon_1}\omega, \\
&\tilde\phi=\frac{\varepsilon_2^2}{\varepsilon_1^2}\phi+\varepsilon_4c_pp^\kappa
  -\frac{\varepsilon_2}{\varepsilon_1^2}\boldsymbol\beta_{tt}(t)\cdot O\mathbf x+\alpha(t), \quad
 \tilde T=\frac{\varepsilon_2^2}{\varepsilon_1^2}T-\varepsilon_4p^\kappa,
\end{split}
\end{gather*}
where $\varepsilon_0$, \dots, $\varepsilon_4$ are arbitrary constants with $\varepsilon_1\ne0$, $\varepsilon_2>0$ and $\varepsilon_3>0$;
$\boldsymbol\beta=(\beta^1,\beta^2)^{\mathsf T}$; the parameters~$\beta^1$, $\beta^2$ and $\alpha$ run through the set of smooth functions of~$t$;
$O$ is an arbitrary $2\times2$ orthogonal matrix.
\end{theorem}

\begin{corollary}
The discrete symmetries of the primitive equations~\eqref{eq:PrimitiveEquations} 
with with $f=0$, $J=0$ and arbitrary values $R>0$ and $\kappa\in(0,1)$ are exhausted,
up to combining with continuous symmetries and with each other, by two involutions,
which are the simultaneous inversion of time and velocity, $(t,u,v,\omega)\to(-t,-u,-v,-\omega)$,
and simultaneous reflections in the $(x,y)$- and $(u,v)$-planes, $(x,y,u,v)\to(-x,y,-u,v)$.
\end{corollary}

\section{Exact solutions}\label{sec:SolutionsPrimitiveEquations}

Finding the transformation~\eqref{eq:PrincipalTransformationPrimitiveEquations} has two more immediate benefits. It allows one to take arbitrary exact solutions of the primitive equations in a resting reference frame to exact solutions of the primitive equations in a constantly rotating reference frame and vice versa. This transformation is also important because it enables one to carry out Lie reductions using the simplified Lie invariance algebra $\mathfrak g_0$, which is spanned by the vector fields~\eqref{eq:MaximalLieInvarianceAlgebraPrimitiveEquationsFzero}, and then to extend the solutions obtained to the rotating case. Examples for both of the above usages are presented in this section.

Physically, the simple solution of the isentropic ($J=0$) primitive equations in a resting reference frame ($f=0$),
\[
 u=u_0(p),\quad v=v_0(p),\quad \omega=0,\quad \phi=\phi(p),\quad T=T(p),
\]
where the relation between~$T$ and~$p$ is given via the hydrostatic equation, describes a stably stratified atmosphere with a horizontally homogeneous horizontal wind field, a vanishing vertical velocity and horizontally homogeneous fields of geopotential and temperature. The inverse of the transformation~\eqref{eq:PrincipalTransformationPrimitiveEquations} takes this solution to
\begin{gather*}
\tilde u = \cos\left(\frac f2t\right)u_0(p)+\sin\left(\frac f2t\right)v_0(p)+ \frac f2y,\\
\tilde v =-\sin\left(\frac f2t\right)u_0(p)+\cos\left(\frac f2t\right)v_0(p)- \frac f2x,\\
\tilde\omega = 0,\quad
\tilde\phi =\phi(p) - \frac{f^2}{8}(x^2+y^2),\quad
\tilde T=T(p),
\end{gather*}
which is a solution of the isentropic primitive equations in a constantly rotating reference frame. The transformed solution is horizontally isotropic in the geopotential, while there is still no vertical velocity. Physically, this means that the effects of a constant rotation cannot lead to vertical motion if the initial vertical velocity is vanishing. The above solution then describes the inertia motion of fluid particles under the action of the Coriolis force, cf.~\cite{ches09Ay} for the corresponding solution of the rotating shallow-water equations. This type of motion can be frequently observed for buoys in the ocean.

To systematically carry out Lie reductions of the primitive equations~\eqref{eq:PrimitiveEquations} with $f=0$ and $J=0$, it is necessary to compute an optimal list of inequivalent subalgebras, which forms the cornerstone of the reduction procedure. We do not aim to establish a complete list of inequivalent subalgebras of dimensions one, two and three here, which for the proper cases would allow reduction of the number of independent variables by one, two or three.
In other words, the corresponding reduced systems would be systems of partial differential equations in two independent variables, systems of ordinary differential equations and systems of algebraic equations, respectively.

Instead, we consider the Lie reduction with respect to the subalgebra 
\[\mathfrak s=\big\langle\XX(\boldsymbol\gamma)+a_1\SSS, \XX(\boldsymbol\sigma)+a_2\SSS\big\rangle,\]
where $a_1$ and $a_2$ are arbitrary constants,
the smooth vector-valued functions $\boldsymbol\gamma=(\gamma^1,\gamma^2)^{\mathsf T}$ and $\boldsymbol\sigma=(\sigma^1,\sigma^2)^{\mathsf T}$ of~$t$ are linearly independent,
and $\boldsymbol\gamma_{tt}\cdot\boldsymbol\sigma-\boldsymbol\sigma_{tt}\cdot\boldsymbol\gamma=0$.
Note that in this case the vector fields $\XX(\boldsymbol\gamma)+a_1\SSS$ and $\XX(\boldsymbol\sigma)+a_2\SSS$ commute and span a proper subalgebra that is suitable for Lie reduction. Previous experience shows that the algebra~$\mathfrak s$ can be included in an optimal list of two-dimensional subalgebras of the algebra~$\mathfrak g_0$, see the corresponding results for the vorticity equation~\cite{bihl09Ay}, the Euler equations~\cite{popo96Ay,popo00Ay}, the Navier--Stokes equations~\cite{fush94AyI,fush94AyII} and the magneto-hydrodynamic equations~\cite{popo97Ay}. An appropriate reduction ansatz constructed with the algebra~$\mathfrak s$~is
\begin{gather*}
\vv = \hat\vv + \frac{\boldsymbol\sigma^\bot\cdot\xx}\delta \boldsymbol\gamma_t-\frac{\boldsymbol\gamma^\bot\cdot\xx}\delta \boldsymbol\sigma_t,\\
\omega = \hat\omega,\\[-.5ex]
\phi = \hat\phi + c_p\frac{p^\kappa}\delta(a_1\boldsymbol\sigma^\bot-a_2\boldsymbol\gamma^\bot)\cdot\xx
- \frac{\boldsymbol\sigma^\bot\cdot\xx}{2\delta}\boldsymbol\gamma_{tt}\cdot\xx-\frac{\boldsymbol\gamma^\bot\cdot\xx}{2\delta}\boldsymbol\sigma_{tt}\cdot\xx,\\[1ex]
T= p^{\kappa}\hat T + \frac{p^\kappa}\delta(a_1\boldsymbol\sigma^\bot-a_2\boldsymbol\gamma^\bot)\cdot\xx,
\end{gather*}
where $\boldsymbol\gamma^\bot:=(\gamma^2,-\gamma^1)$, $\boldsymbol\sigma^\bot:=(\sigma^2,-\sigma^1)$,
$\delta:=\gamma^1\sigma^2-\gamma^2\sigma^1=\boldsymbol\gamma\cdot\boldsymbol\sigma^\bot=-\boldsymbol\gamma^\bot\cdot\boldsymbol\sigma\ne0$
(the function~$\delta$ can be assumed to be positive up to simultaneously alternating signs, e.g., of~$\boldsymbol\gamma^1$ and~$\boldsymbol\gamma^2$),
and quantities with hat depend on the invariant independent variables~$t$ and~$p$.
By the way, for each vector-valued function $\boldsymbol\beta=(\beta^1,\beta^2)^{\mathsf T}$ one has the representation
\[
\boldsymbol\beta
= \frac{\boldsymbol\sigma^\bot\cdot\boldsymbol\beta}\delta \boldsymbol\gamma     -\frac{\boldsymbol\gamma^\bot\cdot\boldsymbol\beta}\delta \boldsymbol\sigma
=-\frac{\boldsymbol\sigma     \cdot\boldsymbol\beta}\delta \boldsymbol\gamma^\bot+\frac{\boldsymbol\gamma     \cdot\boldsymbol\beta}\delta \boldsymbol\sigma^\bot.
\]
The above ansatz reduces the primitive equations~\eqref{eq:PrimitiveEquations} with $f=0$ and $J=0$ to
\begin{subequations}\label{eq:PrimitiveEquationsReducedUsingXX}
\begin{gather}
 \hat\vv_t + \hat\omega\hat\vv_p + \frac{\boldsymbol\sigma^\bot\cdot\hat\vv}\delta \boldsymbol\gamma_t-\frac{\boldsymbol\gamma^\bot\cdot\hat\vv}\delta \boldsymbol\sigma_t
    +c_p\frac{p^\kappa}\delta(a_1\boldsymbol\sigma^\bot-a_2\boldsymbol\gamma^\bot) =0,                            \label{eq:PrimitiveEquationsReducedUsingXX1} \\
 \hat\phi_p + Rp^{\kappa-1}\hat T = 0,                                                                            \label{eq:PrimitiveEquationsReducedUsingXX2} \\
 \frac{\delta_t}\delta+\hat\omega_p = 0,                                                                          \label{eq:PrimitiveEquationsReducedUsingXX3} \\
 \hat T_t + \hat\omega\hat T_p + \frac1\delta(a_1\boldsymbol\sigma^\bot-a_2\boldsymbol\gamma^\bot)\cdot\hat\vv=0. \label{eq:PrimitiveEquationsReducedUsingXX4}
\end{gather}
\end{subequations}
Upon integrating the reduced continuity equation~\eqref{eq:PrimitiveEquationsReducedUsingXX3} to yield
\[\hat \omega = \delta_t\delta^{-1}p+\chi(t),\]
it is clear that the above system is reduced to a linear system of four (1+1)-dimensional first-order partial differential equations,
which can be solved in the following way.
We make the change of dependent variables $\hat\vv=A\check\vv$,
where the $2\times 2$ nondegenerate matrix-valued function $A=A(t)$ is chosen as a solution of the equation $A_t-HA=0$
with the $2\times 2$ matrix-valued function $H=H(t)$ defined by
$H\xx=\delta^{-1}(\boldsymbol\sigma^\bot\cdot\xx)\boldsymbol\gamma_t-\delta^{-1}(\boldsymbol\gamma^\bot\cdot\xx)\boldsymbol\sigma_t$.
Let $A^{-1}$ denote the inverse of the matrix~$A$.
Then the first two equations~\eqref{eq:PrimitiveEquationsReducedUsingXX1} reduce to
\begin{equation}\label{eq:PrimitiveEqsSystemforCheckV}
\check\vv_t + \hat\omega\check \vv_p + c_p\frac{p^\kappa}\delta A^{-1}(a_1\boldsymbol\sigma^\bot-a_2\boldsymbol\gamma^\bot)=0,
\end{equation}
which is an inhomogeneous system of two decoupled linear partial differential equations.
The change of the independent variables
\[
\tau=t,\quad \xi = \frac p{\delta(t)} -\theta(t)\quad\mbox{with}\quad \theta(t)=\int_{t_0}^t\frac{\chi(t')}{\delta(t')}\,\mathrm dt'
\]
maps the system~\eqref{eq:PrimitiveEqsSystemforCheckV} to the system of trivial ordinary differential equations
with the independent variable~$\tau$, where $\xi$ plays the role of parameter.
The general solution of the latter system can be found by quadratures.
This gives the following expression for~$\hat\vv$:
\[
\hat\vv(t,p) = A(t)\vv^0(\xi) - c_pA(t)\int_{t_0}^t
(\delta(\tau))^{\kappa-1}\big(\xi+\theta(\tau)\big)^\kappa
A^{-1}(\tau)\big(a_1\boldsymbol\sigma^\bot(\tau)-a_2\boldsymbol\gamma^\bot(\tau)\big)\,\mathrm d\tau,
\]
where $\vv^0$ is an arbitrary two-dimensional vector-valued smooth functions of~$\xi$.

In order to solve the equation~\eqref{eq:PrimitiveEquationsReducedUsingXX4},
we substitute the obtained expression for~$\hat\vv$ into it,
switch again to the variables~$(\tau,\xi)$ and integrate with respect to~$\tau$.
As a result, we have
\[
\hat T(t,p)=T^0(\xi)-\int_{t_0}^t \frac{a_1\boldsymbol\sigma^\bot(\tau)-a_2\boldsymbol\gamma^\bot(\tau)}{\delta(\tau)}\cdot
\hat\vv\Big(\tau,\delta(\tau)\big(\xi+\theta(\tau)\big)\Big)
\,\mathrm d\tau,
\]
where $T^0$ is an arbitrary smooth function of~$\xi$.

The last step of solving the system~\eqref{eq:PrimitiveEquationsReducedUsingXX} is
the integration of the equation~\eqref{eq:PrimitiveEquationsReducedUsingXX2} with respect to~$p$,
which gives
\[
\hat\phi=\phi^0(t)-R\int_{p_0}^p \tilde p^{\kappa-1}\hat T(t,\tilde p)\,\mathrm d\tilde p,
\]
where $\phi^0$ is an arbitrary smooth function of~$t$, which can be neglected 
in view of the fact that the primitive equations~\eqref{eq:PrimitiveEquations} admits 
gaugings of the geopotential, which depend on~$t$, as their Lie symmetries. 

Substituting the expressions derived for the values with hats into the ansatz,
we obtain the entire family of $\mathfrak s$-invariant solutions of the primitive equations~\eqref{eq:PrimitiveEquations}.

\section[On equivalence transformations within the class of the primitive equations]{On equivalence transformations within the class\\ of the primitive equations}\label{sec:OnEquivTrans}

Given a class~$\mathcal L$ of differential equations, 
it is common to consider only its usual equivalence transformations~\cite{ovsi82a} 
whose components for the independent and dependent variables do not depend on the arbitrary elements of the class~$\mathcal L$. 
Moreover, the infinitesimal method is usually applied for computing such transformations~\cite{akha1991a,ovsi82a}. 
This leads to the usual equivalence algebra of~$\mathcal L$ 
that consists of the infinitesimal generators of one-parameter local groups of equivalence transformations of~$\mathcal L$. 

Recall that, after re-denoting $\kappa:=R/c_p$ and $\hat J:=J/c_p$, the class of systems of the form~\eqref{eq:PrimitiveEquations} 
with arbitrary-element tuple $(f,R,\kappa,\hat J)$ is referred to as the class~\eqref{eq:PrimitiveEquations}. 
The parameters~$f$, $R$, and $\kappa$ are arbitrary constants with $R>0$ and $0<\kappa<1$, 
and the parameter $J$ is an arbitrary sufficiently smooth function of~$(t,x,y,z)$.

\begin{theorem}
The usual equivalence algebra~$\mathfrak g^\sim$ of the class~\eqref{eq:PrimitiveEquations} 
with the arbitrary-element tuple $(f,R,\kappa,\hat J)$ is spanned by the vector fields
\begin{gather*}
\begin{split}
&t\p_t-u\p_u-v\p_v-\omega\p_\omega-2\phi\p_\phi-2T\p_T-f\p_f-3\hat J\p_{\hat J},\\
 &x\p_x+y\p_y+u\p_u+v\p_v+2\phi\p_\phi+2T\p_T+2\hat J\p_{\hat J},\quad
  p\p_p+\omega\p_\omega,\quad 
  T\p_T-R\p_R+\hat J\p_{\hat J},\\
 &\p_t,\quad -y\p_x+x\p_y-v\p_u+u\p_v,\quad \p_x,\quad \p_y,\quad
  \alpha\p_\phi,
\end{split}
\end{gather*}
where the parameter function~$\alpha$ runs through the set of smooth functions depending on $t$.
\end{theorem}

It is surprising that a proper computation of point equivalences within the class~\eqref{eq:PrimitiveEquations} 
is hinted by the study presented in Section~\ref{sec:SymmetriesPrimitiveEquations}. 
Theorem~\ref{thm:ReductionOfPrimitiveEquationsToReferenceFrameAtRest} 
and the form of the vectors fields~$\DD_1$ and~$\SSS$ from the algebra~$\mathfrak g_f$, 
which are presented in Eq.~\eqref{eq:MaximalLieInvarianceAlgebraPrimitiveEquationsF}, show 
that for the class~\eqref{eq:PrimitiveEquations} it is in fact necessary 
to consider its generalized equivalence group rather than its usual equivalence group. 
The notion of generalized equivalence group was introduced in~\cite{mele1996a}, 
see also \cite{opan2017a,opan2019b,opan2020b} and references therein for further developments, 
including the notion of effective generalized equivalence group.
For elements of the generalized equivalence group of a class~$\mathcal L$ of (systems of) differential equations, 
which are called generalized equivalence transformations of this class,
it is allowed that their components for independent and dependent variables 
can depend on the arbitrary elements of~$\mathcal L$. 
The infinitesimal counterpart of the generalized equivalence group of the class~$\mathcal L$ 
is called the generalized equivalence algebra of this class. 
Generalized equivalence algebras are computed similarly to usual ones using the infinitesimal equivalence criterion. 

\begin{theorem}
The generalized equivalence algebra~$\mathfrak g^\sim_{\rm gen}$ of the class~\eqref{eq:PrimitiveEquations} 
with the arbitrary-element tuple $(f,R,\kappa,\hat J)$ consists of the vector fields 
that are sums of the vector fields of the form
\begin{gather*}
\begin{split}
&\zeta^1\big(t\p_t-u\p_u-v\p_v-\omega\p_\omega-2\phi\p_\phi-2T\p_T-f\p_f-3\hat J\p_{\hat J}\big),\\
&\zeta^2\big(x\p_x+y\p_y+u\p_u+v\p_v+2\phi\p_\phi+2T\p_T+2\hat J\p_{\hat J}\big),\quad 
 \zeta^3\big(p\p_p+\omega\p_\omega\big),\\
&\zeta^4\big(T\p_T-R\p_R+\hat J\p_{\hat J}\big),\quad 
 \zeta^5\p_t,\quad 
 \zeta^6\big(-y\p_x+x\p_y-v\p_u+u\p_v\big),\\
&\zeta^7\big(-ty\p_x+tx\p_y-(tv+y)\p_u+(tu+x)\p_v+\tfrac12f(x^2+y^2)\p_\phi-2\p_f\big),\quad 
\\
&\boldsymbol\gamma\cdot\p_{\mathbf x}+\boldsymbol\gamma_t\cdot\p_{\mathbf v}-(\boldsymbol\gamma_{tt}\cdot\mathbf x+f(\gamma^1_ty-\gamma^2_tx))\p_\phi, \quad 
 \alpha\p_\phi,\quad
 p^\kappa\big(\kappa\beta\p_T-R\beta\p_\phi+\kappa\beta_t\p_{\hat J}\big),
\end{split}
\end{gather*}
where the parameter function~$\alpha$, $\beta$, $\gamma^1$ and $\gamma^2$ 
run through the set of smooth functions depending on $(t,f,R,\kappa)$, 
and the parameter function~$\zeta^1$, \dots, $\zeta^7$
run through the set of smooth functions depending on $(f,R,\kappa)$. 
\end{theorem}

Since systems from the class~\eqref{eq:PrimitiveEquations} have 
a number of independent variables, unknown functions and equations, 
and all these equations are of order one, 
it is too difficult to construct the usual and generalized equivalence (pseudo)groups~$G^\sim$ and~$G^\sim_{\rm gen}$ 
of the class~\eqref{eq:PrimitiveEquations} by the direct method. 
At the same time, these groups can be computed by the algebraic method
similarly to the computation of the complete point symmetry (pseudo)group~$G_0$ of the primitive equations~\eqref{eq:PrimitiveEquations} 
with $f=0$, $J=0$ and arbitrary values $R>0$ and $\kappa\in(0,1)$ in Section~\ref{sec:ompletePointSymmetryGroupPrimitiveEquations}.

In view of presence of the constant arbitrary elements $f$, $R$ and~$\kappa$, 
there is definitely a dependence of the parameters of the generalized equivalence algebra~$\mathfrak g^\sim_{\rm gen}$ 
and the generalized equivalence group~$G^\sim_{\rm gen}$ on class arbitrary elements 
that is needless for generating admissible transformations between systems from the class~\eqref{eq:PrimitiveEquations}. 
This is why the study of effective generalized equivalence groups is relevant for the class~\eqref{eq:PrimitiveEquations} 
but such groups can be too difficult for finding 
even in the case of classes of single (1+1)-dimensional partial differential equations and, moreover, they have unusual properties. 

To avoid working with generalized admissible transformations, 
we can split the class~\eqref{eq:PrimitiveEquations} into the subclasses singled out via fixing values of $(f,R,\kappa)$ 
and gauging the constants~$f$ and~$R$ to~$0$ and~$1$ by the transformation~\eqref{eq:PrincipalTransformationPrimitiveEquations} 
with this value of~$f$ and by a usual equivalence transformation of scaling $(T,R,\hat J)$, respectively. 
The maximal Lie invariance algebra of a system of the form~\eqref{eq:PrimitiveEquations} 
with $(f,R)=(0,1)$ and $\kappa\in(0,1)$ is constituted by the vector fields 
\[
Q=c_1\DD_1+c_2\DD_2+c_3\DD_3+c_4\PP+c_5\JJ+\hat\SSS(\beta)+\XX(\boldsymbol\gamma)+\ZZ(\alpha),
\]
where we use the notation of Eq.~\eqref{eq:MaximalLieInvarianceAlgebraPrimitiveEquationsFzero}, 
$\hat\SSS(\beta):=\beta(t)p^\kappa(\p_\phi-\kappa\p_T)$,  
$c_1$, \dots, $c_5$ are arbitrary constants 
and $\alpha$, $\beta$, $\gamma^1$ and~$\gamma^2$ are arbitrary smooth functions of~$t$ 
that satisfy the classifying equation 
\begin{gather}\label{eq:ClassifyingConditionForPrimitiveEquationsWithR1f0}
\begin{split}
&(c_1t+c_4)\hat J_t+(c_2x-c_5y+\gamma^1)\hat J_x+(c_2y+c_5x+\gamma^2)\hat J_y+c_3p\hat J_p\\
&\qquad{}=(2c_2-3c_1)\hat J-\kappa\beta_tp^\kappa.
\end{split}
\end{gather}
In other words, for any system of the form~\eqref{eq:PrimitiveEquations} with $(f,R)=(0,1)$ and $\kappa\in(0,1)$, 
its maximal Lie invariance algebra is contained in the algebra 
\[
\mathfrak g_\spanindex=\big\langle\DD_1,\DD_2,\DD_3,\PP,\JJ,\hat\SSS(\beta),\XX(\boldsymbol\gamma),\ZZ(\alpha)\big\rangle,
\]
where the parameters~$\alpha$, $\beta$, $\gamma^1$ and~$\gamma^2$ run through the set of arbitrary smooth functions of~$t$.
Thus, the problem of group classification of the class~\eqref{eq:PrimitiveEquations}
reduces to the classification of appropriate subalgebras of~$\mathfrak g_\spanindex$.  
A subalgebra~$\mathfrak s$ of~$\mathfrak g_\spanindex$ is called \emph{appropriate} 
if there exists a value of the parameter-function~$J$ such that 
the system of the primitive equations~\eqref{eq:PrimitiveEquations} with this value of~$J$ 
(as well as with the fixed values $(f,R)=(0,1)$ and $\kappa\in(0,1)$) admits~$\mathfrak s$ 
as its maximal Lie invariance algebra. 
This definition can be interpreted from the computational point of view. 
Substituting the components of each vector field~$Q$ from~$\mathfrak s$ 
into the classifying equation~\eqref{eq:ClassifyingConditionForPrimitiveEquationsWithR1f0}
gives an equation with respect to~$J$. 
Let $\mathsf S$ denotes the system of all such equations for~$Q$ running through~$\mathfrak s$. 
We call~$\mathsf S$ the system associated with the subalgebra~$\mathfrak s$. 
A number of independent equations in~$\mathsf S$ is necessarily finite. 
The subalgebra~$\mathfrak s$ is appropriate if and only if
the system~$\mathsf S$ associated with~$\mathfrak s$ is consistent with respect to~$J$ 
and this subalgebra is maximal among the subalgebras of~$\mathfrak g_\spanindex$ 
for which the solutions sets of the associated systems 
coincide with the solutions set of the system~$\mathsf S$. 
Thus, the algebra~$\mathfrak g_0$ is an appropriate subalgebra of~$\mathfrak g_\spanindex$, 
which is singled out by the constraint $\beta_t=0$ and the corresponding value of~$J$ is $J=0$. 
A complication for the group classification of the class~\eqref{eq:PrimitiveEquations} 
is that among the appropriate subalgebras of the algebra~$\mathfrak g_\spanindex$
there are both finite-dimensional and infinite-dimensional algebras.

\section{Conclusion}\label{sec:ConclusionPrimitiveEquations}

The present paper is devoted to an investigation of transformational properties of the primitive equations. 
We have found a point transformation that allows us to cancel the effects of a constant Coriolis force in the primitive equations. 
This transformation might be relevant for application of the primitive equations on the $f$-plane, such as in studies of land-sea breezes.

In practice, the primitive equations~\eqref{eq:PrimitiveEquations} 
as presented in Section~\ref{sec:PrimitiveEquations} are forced and damped by external mechanisms, 
such as external heating (including $J$), phase transitions of water (including an equation for moisture) 
and bottom friction (including friction in the momentum equations). 
The presence of these additional mechanisms substantially narrows the number of admitted Lie symmetries. 
It has been discussed in Section~\ref{sec:PrimitiveEquations} that for arbitrary~$J$ depending on $(t,x,y,p)$ 
only the vector fields~$\ZZ(\alpha)$ and~$\SSS$ are admitted by the system~\eqref{eq:PrimitiveEquations} as Lie symmetries. 
Symmetry breaking due to external forcing terms thus substantially hinders the applicability of symmetry methods. 
On the other hand, by omitting these external influences 
it is possible to arrive at a system that has a wide maximal Lie invariance algebra 
and is therefore accessible to the machinery of group analysis. 
Eventually, results derived for the simplified equations can be extended to the complete system. 
In this manner, we have shown
that the same transformation~\eqref{eq:PrincipalTransformationPrimitiveEquations} 
that maps the rotating primitive equations to the non-rotating ones in the isentropic case 
extends trivially also to the non-isentropic case, 
assuming that $\tilde J(\tilde t,\tilde x,\tilde y,\tilde p)=J(t,x,y,p)$.

Analogously, it can be checked that the transformation~\eqref{eq:PrincipalTransformationPrimitiveEquations} also maps the \emph{dissipative} primitive equations in a constantly rotating reference frame to the dissipative primitive equations in a resting reference frame. This means that attaching classical friction, $\nu\Delta \vv$, to the right-hand side of the momentum equations in~\eqref{eq:PrimitiveEquations} does not require to modify transformation~\eqref{eq:PrincipalTransformationPrimitiveEquations} in order to set~$f$ to zero.

In Section~\ref{sec:PrimitiveEquations}, we have discussed an application of the model of primitive equations for the choice $f=\const$, i.e., on the $f$-plane. For domains extending farther in North--South direction, the latitudinal variation of the Coriolis parameter becomes relevant. The next order of approximation for~$f$ in a Cartesian plane is a linear Taylor polynomial of the form $f=f_0+\beta y$ with $\beta=\const$. It can be checked that in this case, the primitive equations~\eqref{eq:PrimitiveEquations} only admit a six-parameter maximal Lie invariance group which therefore cannot be isomorphic to the maximal Lie invariance algebra~$\mathfrak g_0$ computed in Section~\ref{sec:SymmetriesPrimitiveEquations}. This at once implies that there cannot exist a point transformation which maps the case $f=f_0+\beta y$ to the primitive equations in a resting reference frame.

In Section~\ref{sec:ompletePointSymmetryGroupPrimitiveEquations}, we have established another main result of this paper by computing the complete point symmetry group of the primitive equations using the algebraic method. This computation was rather elaborate due to the multidimensional spaces of both independent and dependent variables of the primitive equations. It was necessary to establish a suitable set of megaideals, finding of which crucially relied on the iterative use of Proposition~1 from~\cite{card12Ay}. Without this set of megaideals it would have been overly difficult to simplify the determining equations for point symmetry transformations of the primitive equations enough to enable their direct integration. This example thus shows the power of the algebraic method for finding complete point symmetry groups for large systems of nonlinear partial differential equations admitting infinite-dimensional Lie pseudogroups, which would be challenging with the conventional direct method.

In Section~\ref{sec:SolutionsPrimitiveEquations}, we have shortly discussed the construction of exact solutions of the primitive equations with rotation. 
The transformation~\eqref{eq:PrincipalTransformationPrimitiveEquations} 
allows one to carry over solutions of the nonrotating equations to the equations in a rotating reference frame. 
This is important from the point of view of exact solutions that can be obtained by Lie reduction, 
as the vector fields from the algebra~$\mathfrak g_0$ are considerable simpler than those from the algebra~$\mathfrak g_f$. 
We have derived exact solutions of the primitive equations that arise from a completely integrable case of Lie reduction. 
The case considered is certainly the most important example of reduction with respect to two-dimensional subalgebras. 
We should also like to stress that upon constructing the optimal lists of inequivalent subalgebras, 
a considerable fraction of these lists will not be suitable for Lie reduction. 
In particular, all algebras including $\ZZ(\alpha)$ for some $\alpha$, $\langle\SSS\rangle$ 
or a combination of these two vector fields will not allow one to find a reduction ansatz. 
Moreover, a number of three-dimensional subalgebras will also not be needed for reduction 
even if they are appropriate for it. 
More precisely, the Lie reduction using any algebra containing a subalgebra 
$\mathfrak s=\big\langle\XX(\boldsymbol{\gamma})+c_1\SSS, \XX(\boldsymbol{\sigma})+c_2\SSS\big\rangle$ for some constants~$c_1$ and~$c_2$ and some linearly independent two-dimensional vector-valued functions $\boldsymbol{\gamma}$ and $\boldsymbol{\sigma}$ of~$t$ with $\boldsymbol{\gamma}_{tt}\cdot\boldsymbol{\sigma}-\boldsymbol{\sigma}_{tt}\cdot\boldsymbol{\gamma}=0$ is not required 
since it results only in a subfamily of the family of $\mathfrak s$-invariant solutions, 
an this entire family is explicitly constructed in quadratures in Section~\ref{sec:SolutionsPrimitiveEquations}.  

In view of the remarks of the previous paragraph, despite we have not systematically followed the steps of group-invariant reduction, the results obtained are in a certain sense a substantial part of the exact solutions of the primitive equations that can be found by Lie reduction. A more detailed exposure of the group analysis of the system of primitive equations will be presented elsewhere. 
In the course of finding exact solutions of the primitive equations, the method of group foliations and constructing differential constraints for the primitive equations using differential invariants of their maximal Lie invariant algebra may be relevant. In~\cite{golo2004a,golo2004d}, these tools were applied to the Euler and Navier--Stokes equations for  incompressible fluids. The maximal Lie invariance algebras of both these systems contain infinite-dimensional ideals of the same formal form as $\big\langle\XX(\boldsymbol\gamma),\ZZ(\alpha)\big\rangle$.

The consideration of Section~\ref{sec:OnEquivTrans} creates a basis for a deeper group analysis of the primitive equations.  
At the same time, if we generalize the form of the arbitrary element~$J$ 
via allowing it to depend on other arguments 
like the temperature~$T$ or the vertical velocity component~$\omega$ 
or even to be a differential function, 
then this can leads to changing the corresponding usual and generalized equivalence algebras~$\mathfrak g^\sim$ and~$\mathfrak g^\sim_{\rm gen}$ 
and the corresponding algebra~$\mathfrak g_\spanindex$ underlying the group classification of the primitive equations.

\section*{Acknowledgements}

The authors thank the anonymous reviewers for their valuable comments and suggestions.
The authors are also grateful to Alexander Bihlo and Galyna Popovych for helpful discussions. 
This research was supported by the Austrian Science Fund (FWF), projects P25064 and P28770.

\footnotesize


\end{document}